\documentclass[11pt,a4paper]{article}

\usepackage[T1]{fontenc}
\usepackage[utf8]{inputenc}
\usepackage{soul}
\usepackage{amssymb,amsmath,amsthm,amsfonts}    % math stuff

\usepackage{bm}                                 % bold math
\usepackage{mathtools}                          % tools to create operators and delimiters
\usepackage{stmaryrd}                           % additional math symbols
\usepackage{breqn}                              % automatic line breaking in equations (use wisely)

\usepackage[textfont=footnotesize,labelfont={footnotesize,bf}]{caption}
\usepackage{subcaption}
\usepackage{booktabs}

\usepackage{tikz}
\usepackage{pgfplots}

\usepgfplotslibrary{colorbrewer,colormaps}
\usepgfplotslibrary[fillbetween]
\usepackage{pgfplotstable}
\usetikzlibrary{shapes.geometric}
\usetikzlibrary{external, patterns}
\usetikzlibrary{angles, quotes, decorations.text}

\usepackage[inline]{enumitem}
\usepackage{graphicx}
\usepackage{graphbox} 
\usepackage[top=30mm,right=30mm,left=30mm,bottom=30mm]{geometry}
\usepackage[colorlinks=true,linkcolor=blue,urlcolor=blue]{hyperref}
\usepackage[affil-it,auth-sc]{authblk}          % handle affiliations
\usepackage[colorinlistoftodos,textwidth=25mm,textsize=footnotesize]{todonotes}
\usepackage{lipsum}
\usepackage{multirow}
\usepackage{multicol}
\usepackage{dsfont}
\usepackage{comment}
\usepackage{tabularx}

\usepackage[ backend=biber, style=numeric, citestyle=numeric-comp, maxbibnames=99, minbibnames=99, maxcitenames=2, mincitenames=1, giveninits=true, sorting=none, sortlocale=auto, natbib=true, url=false,isbn=false,
doi=true,eprint=true ]{biblatex}
\newcommand{\ba}{\bm a}

\newcommand{\be}{\bm e}

\newcommand{\bq}{\bm q}

\newcommand{\bK}{\bm K}

\newcommand{\bQ}{\bm Q}

\newcommand{\mU}{\mathcal U}

\graphicspath{{./}{./figures/}}                 % path to figures folder

\title{Homogenization of elastic grids containing rigid elements}

\author[1]{Luca Viviani}
\author[1]{Davide Bigoni\footnote{Corresponding author: e-mail: \href{mailto:bigoni@ing.unitn.it}{bigoni@ing.unitn.it}; phone: +39\,0461\,282507.}}
\author[1]{Andrea Piccolroaz}

\affil[1]{Department of Civil, Environmental, and Mechanical Engineering, University of Trento, Trento, Italy}

\date{Dedicated to Professor Alan Needleman on the occasion of his 80th birthday} 

\begin{document}

\maketitle

\begin{abstract}
\noindent
The inclusion of rigid elements into elastic composites may lead to superior mechanical properties for the equivalent elastic continuum, such as, for instance, extreme auxeticity. To allow full exploitation of these properties, a tool for the homogenization of two-dimensional elastic grids containing rigid elements is developed and tested on elaborate geometries, such as, for instance, Chinese lattices. The rigid elements are assumed to be either jointed with full continuity of displacement or hinged to the elastic rods. It is shown that the two different constraints induce strongly different mechanical characteristics of the equivalent elastic solid. The presented results open the way to the design of architected materials or metamaterials containing both elastic and rigid parts. 
\end{abstract}

\paragraph{Keywords} 
Periodic lattice \textperiodcentered\
Homogenization \textperiodcentered\
Rigid constraint

\section{Introduction}
\label{Sec_introduction}

Homogenization of elastic composites leading to an equivalent continuum is a well-developed theory, a topic subjected to a thorough research effort \cite{born, willis1982elasticity, pontesuq, greame, Kalamkarov,christensen2012mechanics, nemat2013micromechanics}, and represents the crucial tool in the design of architected materials and metamaterials \cite{fleckn}. 
In particular, homogenization techniques for lattices have been addressed for truss structures   \cite{hutchinson2006structural, elsayed2010analysis} and for continuously jointed beams \cite{gibson1982mechanics, masters1996models, arabnejad2013mechanical, vigliotti}, also keeping into account higher-order effects \cite{kumar, sab, ostoja2002lattice, eremeyev2019two}.

It is known that the inclusion of rigid elements inside an elastic composite may yield useful and sometimes even extreme mechanical properties. An example is the hexagonal grid reaching the limit of Poisson's ratio equal to -1, thanks to elements which are flexurally rigid but axially compliant   \cite{rothenburg1991microstructure, milton1992composite}. 
Rigid elements connected through elastic hinges (in other words, truss structures) have been considered in the quest for auxetic behaviour \cite{grima2000auxetic, ishibashi2000microscopic}, or to identify polar elastic materials \cite{vasiliev2002elastic}. 

Despite their interest in the development of metamaterials, a systematic treatment of homogenization for elastic composites containing rigid elements is still lacking and is therefore the subject of the present article. 
In particular, homogenization theory is developed for two-dimensional elastic grids of beams containing rigid parts of arbitrary characteristics, leading to an equivalent elasticity tensor. 

The rigid parts are assumed to be either fully jointed or hinged to the elastic elements. It is shown that the different degree of connection leads to strong differences in the mechanical properties of the equivalent continuum. In particular, closed circuits of rigid elements hinged to each other still permit deformation of the inner elastic beams, becoming impossible when the elements are fully jointed.

Applications of the developed tool reveal several effects related to the presence of rigid inclusions. In particular, it is shown that there are geometries for which the Poisson's ratio of the isotropic equivalent continuum or, more in general, the coefficients of mutual influence of an equivalent triclinic material, 
turn out to be independent of the rigid phase, while in other cases they are not. Moreover, all the examples with fully connected rigid elements show a singularity of 1st-order in the elastic constants when the limit of a completely rigid unit cell is approached. 

Our results may be employed in the design of architected materials in which a rigid phase is exploited to obtain \lq exotic' mechanical properties.

\section{Homogenization of elastic grids with rigid elements}
\label{Sec_model}

Homogenization of two-dimensional lattices of elastic beams, including rigid elements, is developed following the approach presented in \cite{bordiga2019free, bordiga2021dynamics,bordiga2022tensile}. In particular, homogenization is developed for a unit cell composed of elastic flexible and axially deformable beams, jointed together with rigid parts.

\subsection{The mechanics of an elastic beam}

Consider a two-dimensional elastic grid endowed with rigid elements connecting two or more nodes. The periodic structure of this grid is obtained by repeating a unit cell along two directions $\{ \bm{a}_{1}, \bm{a}_{2} \}$. Referring to any of the elastic beams of length $l$, in a local reference system, its in-plane displacement field $\bm{u}$ of \emph{axial} and \emph{transverse} components $u$ and $v$, respectively, is a function of the local variable $s$, ranging between 0 and $l$. The adoption of Euler-Bernoulli beam theory relates the rotation of the cross-section $\theta$ to the derivative of the transverse displacement as $\theta(s) = v'(s)$. The equilibrium equations for the beam and their integrals are
\begin{equation}
\label{eq_solution_general}
    u''(s) = 0, ~\Rightarrow~ u(s) = C_{1} + C_{2} s \, , 
    \quad
    v''''(s) = 0, ~\Rightarrow~ v(s) = D_{1} + D_{2} s + D_{3} s^{2} + D_{4} s^{3} \, ,
\end{equation}
where $C_{i}$ and $D_{i}$ are constants. The nodal displacements of the beam can be collected in the vector
\begin{equation}
\label{eq_nodal_disp}
    \bm{q}= [ u(0), v(0), \theta(0), u(l), v(l), \theta(l) ]^{\mathsf{T}} \, , 
\end{equation}
so that it is possible to represent the displacement field as
\begin{equation}
\label{pera}
    \bm{u}(s)=  
    \begin{bmatrix}
        \displaystyle 1-\frac{s}{l} & 0 & 0 & \displaystyle \frac{s}{l} & 0 & 0 \\[3mm]
        0 & \displaystyle \left(1-\frac{s}{l}\right)^{2}\left(1+2\frac{s}{l}\right) & \displaystyle \left(1-\frac{s}{l}\right)^{2}s & 0 & \displaystyle \left(3-2\frac{s}{l}\right)\frac{s^{2}}{l^{2}} & \displaystyle (s-l)\frac{s^{2}}{l^{2}}
    \end{bmatrix} \,\bm{q} \, .
\end{equation}
The corresponding elastic strain energy is
\begin{equation}
\label{eq_strain_energy}
     \mathcal{U} = \frac{1}{2} \bm{q} \cdot \bm{K} \bm{q} \, ,
\end{equation}
where $\bm{K}$ is the stiffness matrix, which through equation \eqref{pera} results to be
\begin{equation}
    \bm{K} 
    = 
    \begin{bmatrix}
        EA/l & 0 & 0 & -EA/l & 0 & 0 \\
        0 & 12 EI/l^{3} & 6 E I /l^{2}& 0 & -12 EI/l^{3} & 6 E I/l^{2} \\
        0 & 6 EI/l^{2} & 4 EI/l & 0 & -6 EI/l^{2} & 2 EI/l \\
        -EA/l & 0 & 0 & EA/l & 0 & 0 \\
        0 & -12 EI/l^{3} & -6 E I /l^{2}& 0 & 12 EI/l^{3} & -6 E I/l^{2} \\
        0 & 6 EI/l^{2} & 2 EI/l & 0 & -6 EI/l^{2} & 4 EI/l \\
    \end{bmatrix} \, ,
\end{equation}
being $EA$ and $EI$ the axial and flexural stiffnesses of the rod. Notably, this is the stiffness matrix commonly used in the discretization of Euler-Bernoulli beams within the finite element approximation. However, in the present case, the displacement field \eqref{pera} and the corresponding elastic energy \eqref{eq_strain_energy} are exact due to the absence of prestress in the beam and the absence of distributed loading along its axis. Instead, the loading is applied only at the end of the nodes.

\subsection{Modelling of rigid connections}
\label{sec_model_rigid_incl}

\begin{figure}
    \centering
    \includegraphics[width=\textwidth, align=c]{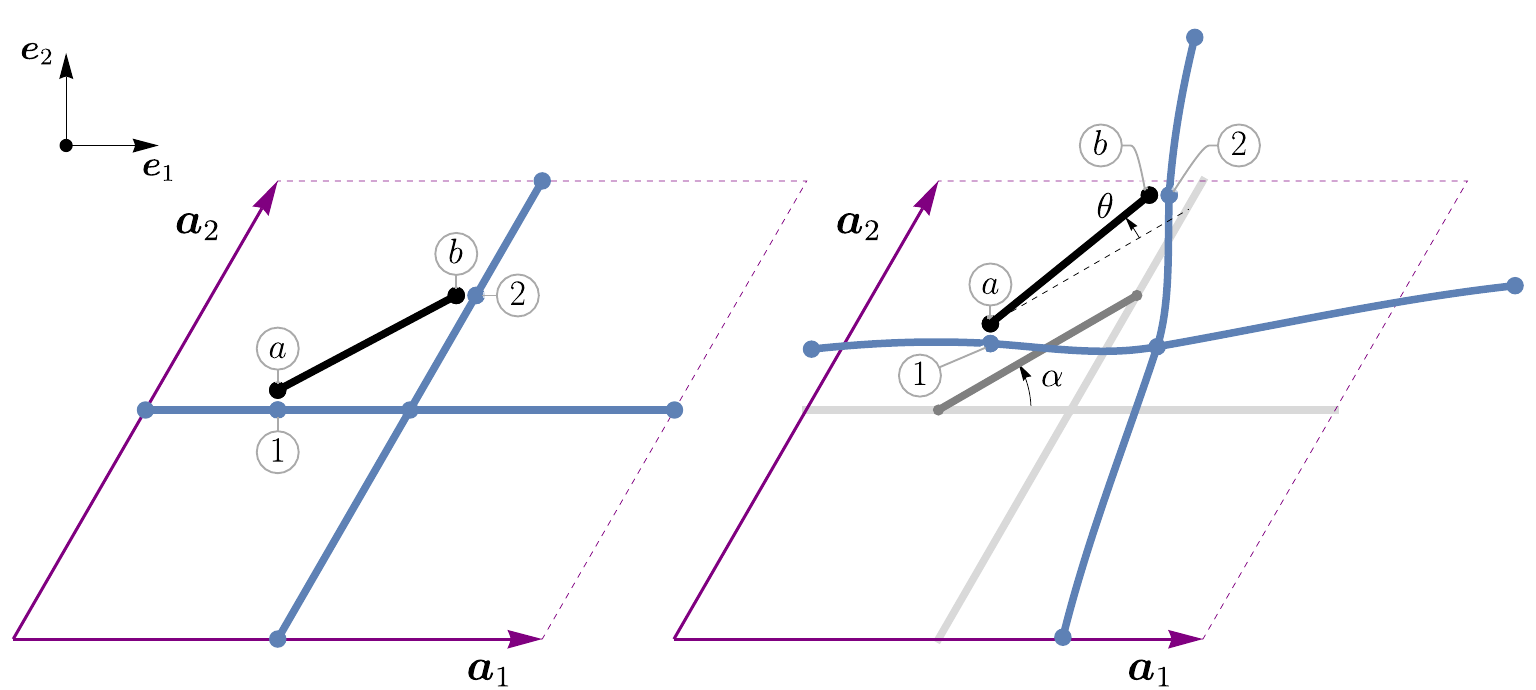}
    \caption{Unit cell composed of Euler-Bernoulli beams and one rigid element connecting the nodes 1-2 in its undeformed (left) and deformed (right) configurations.}
    \label{fig_unitcell_example}
\end{figure}

The presence of rigid elements inside the unit cell introduces a set of constraints, effectively reducing the number of its degrees of freedom. For instance, for the unit cell represented in Fig.~\ref{fig_unitcell_example} (left part), one rigid element links the nodes 1-2, of coordinates $\{x_{1}, \, y_{1}\}$ and $\{ x_{2}, \, y_{2} \}$ in the global $oxy$--reference system, characterized by the two orthogonal unit vectors $\bm{e}_1$ and $\bm{e}_2$. In the following, the homogenization procedure will lead to elasticity tensors given in components in that reference system. 
%
%%%%%%%%%%%%%%%%%%%%%%%%%%%%%%%%%%%%%%%%%%%%%%%%%%%%%%%%%%%%%%%%%%%%%%%%%%%
%%%%%%%%%%%%%%%%%%%%%%%%%%%%%%%%%%%%%%%%%%%%%%%%%%%%%%%%%%%%%%%%%%%%%%%%%%%

% \begin{figure}
% \centering
% \begin{subfigure}[]{\textwidth}
% \centering
% \includegraphics[width=\textwidth, align=c]{figures/cell_rigidbeam.pdf}
% \subcaption[]{}
% \label{fig_lattice_rigid}
% \end{subfigure}
% \\[15pt]
% \begin{subfigure}[]{0.6\textwidth}
% \centering
% \includegraphics[width=\textwidth, align = c]{figures/rigid_beam_motion_2.pdf}
% \subcaption[]{}
% \label{fig_rigidbeam_motion}
% \end{subfigure}
% \caption{(a) Unit cell composed of Euler-Bernoulli beams and one rigid element  connecting the nodes 1-2 in its  \emph{undeformed} (left) and \emph{deformed} (right) configurations. (b) Schematic representation of the linearized displacement performed by the rigid element.  }
% \label{fig_unitcell_example}
% \end{figure}
%%%%%%%%%%%%%%%%%%%%%%%%%%%%%%%%%%%%%%%%%

During a deformation process, as sketched in Fig.~\ref{fig_unitcell_example} (right part), the displacements (of components $u$ and $v$) and rotations ($\theta$) of the ends $(a)$ and $(b)$ of a rigid element of length $\ell$ are related for a finite transformation through
\begin{equation}
    \theta_{(b)} = \theta_{(a)} = \theta \,, \quad
    u_{(b)} = u_{(a)} - \ell [\cos\alpha - \cos(\alpha + \theta)] \,, \quad
    v_{(b)} = v_{(a)} + \ell [ \sin(\alpha + \theta) - \sin\alpha ] \,, 
\end{equation}
which for a small rotation, can be linearized, in agreement with Poisson's theorem, in the form
\begin{equation}
    \theta_{(b)} = \theta_{(a)} =\theta \,, \quad 
    u_{(b)} = u_{(a)} -\theta\,\ell \sin\alpha \,, \quad 
    v_{(b)} = v_{(a)} +\theta \, \ell \cos\alpha \, . 
\end{equation}

How the displacement of the rigid element transmits to a node is influenced by how the connection between the element and the node is realized. Two cases will be analysed: (i.) a perfectly bonded, called \lq clamped', and (ii.) a \lq hinged', connection. 

In the case (i.) of a clamped connection, there is full continuity of the displacement and rotation between the rigid body and the node. On the other hand, in the case (ii.) of a hinged connection, only the displacement remains continuous between the rigid body and the node, while the two rotations are unrelated. 

Considering now two nodes, say, 1 and 2, of the structure, jointed via a rigid element, the four situations listed in Table~\ref{tab_constraint_rigid_beam} can arise. The effect of the rigid constraint is to reduce the degrees of freedom of the unit cell, so that the description of the unit cell is eventually condensed. 
%
%%%%%%%%%%%%%%%%%%%%%%%%%%%%%%%%%%%%%%%%%%%%%%%%%%%%%%%%%%%%%%%%%%%%%%%%%%%
\begin{table}[h!]
    \centering
    \begin{tabular}{@{} l c l @{}} 
        \toprule
        \texttt{Clamped 1 -- Clamped 2} 
        & 
        \hspace{5mm}
        &
        \texttt{Hinged 1 -- Hinged 2} \\
        \cmidrule(r){1-1} \cmidrule(l){3-3}
        \(
            \displaystyle
            \begin{aligned}
            & \theta_{2} = \theta_{1} \notag \\
            & u_{2} = u_{1} - (y_{2} - y_{1}) \theta_{1} \notag \\
            & v_{2} = v_{1} + (x_{2} - x_{1}) \theta_{1} \notag 
            \end{aligned}
        \)
        & 
        &
        \(
            \displaystyle
            \begin{aligned}
            & (u_{2} - u_{1})(x_{2} - x_{1}) = \\ 
            & -(v_{2} - v_{1}) (y_{2} - y_{1}) \notag 
            \end{aligned}
        \)
        \\ 
        \midrule
        \texttt{Clamped 1 -- Hinged 2}
        & 
        \hspace{5mm}
        &  
        \texttt{Hinged 1 -- Clamped 2} \\
        \cmidrule(r){1-1} \cmidrule(l){3-3}
        \(
            \displaystyle
            \begin{aligned}
            & u_{2} = u_{1} - (y_{2} - y_{1}) \theta_{1} \notag \\
            & v_{2} = v_{1} + (x_{2} - x_{1}) \theta_{1} \notag 
            \end{aligned} 
        \)
        &
        &
        \(
            \displaystyle
            \begin{aligned}
            & u_{2} = u_{1} - (y_{2} - y_{1}) \theta_{2} \notag \\
            & v_{2} = v_{1} + (x_{2} - x_{1}) \theta_{2} \notag 
            \end{aligned} 
        \)
        \\
        \bottomrule
    \end{tabular}
    \caption{List of linearized kinematic conditions to be imposed between node 1 (of coordinates $\{x_1,y_1\}$) and node 2 (of coordinates $\{x_2,y_2\}$) when the two nodes are connected with a rigid element.}
    \label{tab_constraint_rigid_beam}
\end{table}
%%%%%%%%%%%%%%%%%%%%%%%%%%%%%%%%%%%%%%%%%%%%%%%%%%%%%%%%%%%%%%%%%%%%%%%%%%%

\subsection{The mechanics of the unit cell}

The equilibrium equations for the unit cell can be obtained through the \emph{principle of virtual work}, by imposing
\begin{equation}
    \delta \mathcal{U} = \bm{f} \cdot \delta \bm{q} \, \qquad \forall \delta \bm{q} \, ,
\end{equation}
where $\delta \mathcal{U}$ is the internal virtual work corresponding to a virtual displacement, and $\bm{f} \cdot \delta \bm{q}$ represents its external counterpart. Being the unit cell formed of $N_{b}$ elastic beams and $N_{r}$ rigid elements, the strain energy is obtained by summing contributions from each elastic rod 
\begin{equation}
  \delta \mathcal{U} = \sum_{k=1}^{N_{b}} \delta \mU_{k} (\bm{C}_{k} \bm{q}) = \bK \bq \cdot \delta\bq \, ,
\end{equation}
where the dependency on the connectivity matrix $\bm{C}_{k}$ is given evidence because it is needed to set the appropriate end conditions for different beams (continuity of rotations or displacements and so on). In fact, it is $\bm{q}_{k}=\bm{C}_{k} \bm{q}$, with $\bm{q}$ denoting all the degrees of freedom of the unit cell.

Excluding the presence of body forces, the unit cell, ideally \lq excised' from the rest of the grid, transmits forces through its boundary. As a consequence, the generalized forces $\bm{f}$ perform virtual work only at the boundary nodes. 

In conclusion, the system of equilibrium equations results to be 
\begin{equation}
\label{eq_equilibrium_incremental}
    \bm{K} \bm{q} = \bm{f} \, ,
\end{equation}
where $\bm{K} =\partial^{2} \mathcal{ U}/ \partial \bm{q}^{2} $ is the stiffness matrix of dimension $3 \, N_{J} \times 3 \, N_{J}$, with $N_{J}$ representing the total number of nodes of the unit cell.

\subsection{Homogenization}
\label{Sec_homogenization}

The homogenization technique is based on the Hill-Mandel theorem \cite{hill1963elastic,mandel1966contribution}, linking the \emph{microscopic} scale of the grid to the \emph{macroscopic} scale of the equivalent continuum. After the introduction of appropriate conditions on the periodicity of the fields in the lattice, the average strain energy density of the continuum is set to be equal to the energy density of the lattice under an equivalent measure of deformation. This is known as macro-homogeneity condition, needed to obtain the constitutive tensor of the two-dimensional continuum equivalent to the discrete structure. 

The generalized displacement vector referred to the $j$-th node in the lattice $\bm{q}^{(j)}=\left[u^{(j)}, v^{(j)}, \theta^{(j)} \right]$, located at $\bm{x}^{(j)}$, is assumed to be given by the sum of periodic functions $\tilde{u}^{(j)}$, $\tilde{v}^{(j)}$, $\tilde{\theta}^{(j)}$ and an affine contribution $\bm{\varepsilon}\, \bm{x}^{(j)}$, modulated by the uniform strain tensor $\bm{\varepsilon}$, the so-called Cauchy--Born kinematics \cite{tadmor2011modeling}, so that
\begin{equation}
    \left[u^{(j)}, v^{(j)}, \theta^{(j)} \right]
    =
    \left[\tilde{u}^{(j)} + \left(\bm{\varepsilon}\, \bm{x}^{(j)}\right)_1, 
    ~
    \tilde{v}^{(j)} + \left(\bm{\varepsilon}\, \bm{x}^{(j)}\right)_2, 
    ~
    \tilde{\theta}^{(j)} \right] .
\end{equation}
Periodicity implies that for every pair of nodes $p$, $q$ and integers $n_1, n_2 \in \{0,1\}$  
\begin{equation}
\label{carote}
    \bm{x}^{(p)} - \bm{x}^{(q)} = n_1\bm{a}_1+n_2\bm{a}_2, 
    \quad
    \Longrightarrow
    \quad
    \tilde{u}^{(p)} = \tilde{u}^{(q)}, 
    \quad
    \tilde{v}^{(p)} = \tilde{v}^{(q)},
    \quad
    \tilde{\theta}^{(p)} = \tilde{\theta}^{(q)}.
\end{equation}
The above conditions are identical to the long-wavelength limit of the Floquet--Bloch conditions used for wave propagation in periodic media. 

\begin{figure}[h!]
    \centering
    \includegraphics[width=0.70\textwidth]{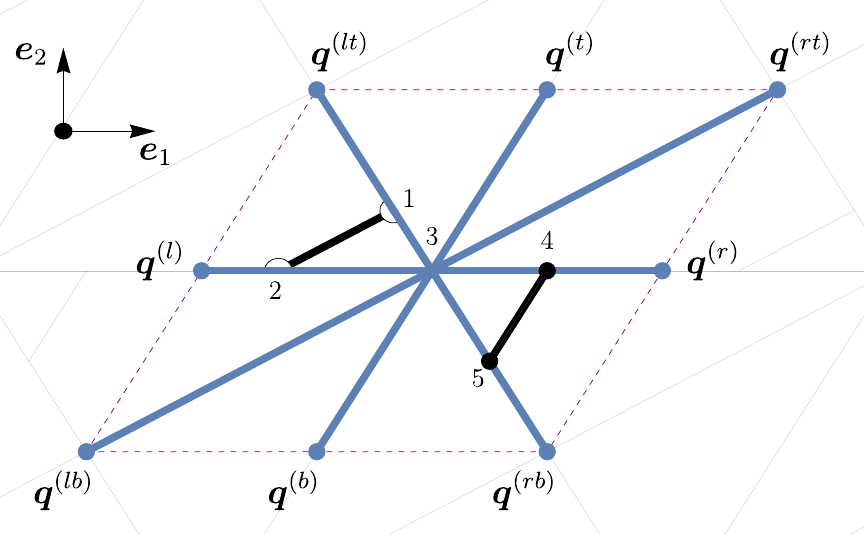}
    \caption{Schematics of a unit cell of parallelepiped-shape, with two internal rigid elements, one connecting the nodes $(4)$ and $(5)$ with full continuity of all generalized displacements, the other connecting nodes (1) and (2) with a hinged contact. The internal nodes are 5 and all their displacement components are collected in the vector $\displaystyle \bm{q}^{(i)}= \{ \bm{q}^{(1)}, \bm{q}^{(2)}, ..., \bm{q}^{(5)} \}$. The displacements of the 8 boundary nodes are gathered as follows: 4 in vectors $\bm{q}^{(lt)}$, $\bm{q}^{(rt)}$, $\bm{q}^{(lb)}$, $\bm{q}^{(rb)}$ pertaining to the corners and 4 in $\bm{q}^{(l)}$, $\bm{q}^{(r)}$, $\bm{q}^{(b)}$, $\bm{q}^{(t)}$, corresponding to the nodes internal to the edges.}
    \label{fig_floquet_bloch_partition}
\end{figure}

Concerning Fig.~\ref{fig_floquet_bloch_partition}, a generic parallelepiped-shaped cell is assumed in which there are $n$ internal nodes (including those connected to a rigid inclusion) and 8 boundary nodes, 4 defining the corners and 4 inside the lateral edges. In terms of generalized displacements, and denoting with a superimposed tilde their periodic part, the above conditions \eqref{carote} specialize to 
\begin{equation}
\label{carotona}
    \tilde{\bm{q}}^{(l)} = \tilde{\bm{q}}^{(r)} \,, \quad
    \tilde{\bm{q}}^{(b)} = \tilde{\bm{q}}^{(t)} \,, \quad
    \tilde{\bm{q}}^{(lb)} = \tilde{\bm{q}}^{(rb)} = \tilde{\bm{q}}^{(rt)} = \tilde{\bm{q}}^{(lt)} \, , 
\end{equation}
where $\bm{q}^{(lt)}$, $\bm{q}^{(rt)}$, $\bm{q}^{(lb)}$, $\bm{q}^{(rb)}$ are the vectors collecting the generalized displacements of the nodes located at the corners, while $\bm{q}^{(l)}$, $\bm{q}^{(r)}$, $\bm{q}^{(b)}$, $\bm{q}^{(t)}$ refer to the nodes internal to the edges of the parallelepiped. The displacement components of internal nodes are all collected in the vector 
\[
    \bm{q}^{(i)} = 
    \{ 
    \underbrace{\bm{q}^{(1)}, \bm{q}^{(2)}, ..., \bm{q}^{(n)}}_{\text{internal nodes}} 
    \} \, ,
\]
which therefore does not only gather the displacements of the node $i$, but all displacements of all internal nodes, $n=5$ in the figure.
%
%%%%%%%%%%%%%%%%%%%%%%%%%%%%%%%%%%%%%%%%%%%%%%%%%%%%%%%%%%%%%%%%%%%%
%%%%%%%%%%%%%%%%%%%%%%%%%%%%%%%%%%%%%%%%%%%%%%%%%%%%%%%%%%%%%%%%%%%%

Finally, it is convenient to gather all the periodic components of generalized displacements in the vector
\begin{equation}
    \tilde{\bm{q}} = [ 
    \tilde{\bm{q}}^{(i)}, \, 
    \tilde{\bm{q}}^{(l)}, \, 
    \tilde{\bm{q}}^{(b)}, \, 
    \tilde{\bm{q}}^{(lb)}, \, 
    \tilde{\bm{q}}^{(r)}, \, 
    \tilde{\bm{q}}^{(t)}, \, 
    \tilde{\bm{q}}^{(rb)}, \, 
    \tilde{\bm{q}}^{(lt)}, \, 
    \tilde{\bm{q}}^{(rt)}
    ]^{\mathsf{T}} \, ,
\end{equation}
so that conditions (\ref{carotona}) allow a condensation of periodic components of generalized displacement into a vector which only contains the independent periodic displacement components $\tilde{\bm{q}}^{*}$ as
\begin{equation}
\label{eq_cauchy_born}
    \tilde{\bm{q}} = \bm{Z}_{0} \tilde{\bm{q}}^{*} \, ,
    \quad \text{ with } \quad
    \bm{Z}_{0} =
    \begin{bmatrix}
    \bm{I} & \bm{0} & \bm{0} & \bm{0} \\
    \bm{0} & \bm{I} & \bm{0} & \bm{0} \\
    \bm{0} & \bm{0} & \bm{I} & \bm{0} \\
    \bm{0} & \bm{0} & \bm{0} & \bm{I} \\
    \bm{0} & \bm{I} & \bm{0} & \bm{0} \\
    \bm{0} & \bm{0} & \bm{I} & \bm{0} \\
    \bm{0} & \bm{0} & \bm{0} & \bm{I} \\
    \bm{0} & \bm{0} & \bm{0} & \bm{I} \\
    \bm{0} & \bm{0} & \bm{0} & \bm{I} \\
    \end{bmatrix} \, ,
    \quad \text{ and } \quad
    \tilde{\bm{q}}^{*} =
    \begin{bmatrix}
    \tilde{\bm{q}}^{(i)} \\ 
    \tilde{\bm{q}}^{(l)} \\ 
    \tilde{\bm{q}}^{(b)} \\
    \tilde{\bm{q}}^{(lb)} 
    \end{bmatrix} \, ,
\end{equation}
where $\bm{I}$ represents the identity matrix in one case of the order enforced by the dimension of the vector $\tilde{\bm{q}}^{(i)}$, while for vectors $\tilde{\bm{q}}^{(l)}$, $\tilde{\bm{q}}^{(b)}$, and $\tilde{\bm{q}}^{(lb)}$ the order is 3.
Note that a more general formulation in which a generic number of nodes is considered inside the edges of the parallelepiped-shaped cell can easily be implemented. 

Finally, the generalized displacements of all nodes forming the unit cell, $\bm{q}$, can be obtained by adding an affine component $\hat{\bm{q}}(\bm{\varepsilon})=[\bm{\varepsilon}\, \bm{x}^{(j)}, \, \bm{0} ]^{\mathsf{T}}$ to a periodic component, in order to describe the kinematics of the unit cell as
\begin{equation}
\label{perotta}
    \bm{q}(\bm{\varepsilon}, \tilde{\bm{q}}^{*}) = \hat{\bm{q}}(\bm{\varepsilon}) + \bm{Z}_{0} \tilde{\bm{q}}^{*} \, .
\end{equation}
A substitution of the representation for generalized displacements of the unit cell, eq.~\eqref{perotta}, into the equilibrium equations \eqref{eq_equilibrium_incremental} and a left multiplication by $\bm{Z}_{0}^{\mathsf{T}}$ leads to 
\begin{equation}
\label{eq_homogenization_1}
    \bm{Z}_{0}^{\mathsf{T}} \bm{K} \bm{Z}_{0} \tilde{\bm{q}}^{*} + 
    \bm{Z}_{0}^{\mathsf{T}} \bm{K} \hat{\bm{q}}(\bm{\varepsilon}) = 
    \bm{Z}_{0}^{\mathsf{T}} \bm{f} \, . 
\end{equation}
The term on the right of eq.~\eqref{eq_homogenization_1} possesses the partition dictated by eq.~\eqref{eq_cauchy_born}, and has to vanish, because anti-periodic forces are generated by periodic displacements,
\begin{equation}
     \bm{Z}_{0}^{\mathsf{T}} \bm{f} = 
     \left[
     \bm{f}^{(i)}, 
     \bm{f}^{(l)} + \bm{f}^{(r)}, 
     \bm{f}^{(b)} + \bm{f}^{(t)}, 
     \bm{f}^{(lb)} + \bm{f}^{(rb)} + \bm{f}^{(lt)} + \bm{f}^{(rt)}
     \right]^{\mathsf{T}} 
     = \bm{0}, 
\end{equation}
so that, in the absence of external forces applied to the grid, it follows  
\begin{equation}
\label{eq_homogenization_x}
    \bm{Z}_{0}^{\mathsf{T}} \bm{K} \bm{Z}_{0} \tilde{\bm{q}}^{*} = 
    - \bm{Z}_{0}^{\mathsf{T}} \bm{K} \hat{\bm{q}}(\bm{\varepsilon}) \, .
\end{equation}
The matrix $\bm{Z}_{0}^{\mathsf{T}} \bm{K} \bm{Z}_{0}$ in eq.~(\ref{eq_homogenization_x}) is always singular whatever unit cell is selected. This follows from $\bm{Z}_{0}^{\mathsf{T}} \bm{K} \bm{Z}_{0}\bm{t} = \bm{0}$, occurring for every vector $\bm{t}$ representing a rigid-body translation, which in the plane is the linear combination of two orthogonal vectors, say, $\bm{t}_1$ and $\bm{t}_2$. Assuming that this translation is the only source of singularity (the null space has dimension 2), Fredholm's alternative theorem states that the solution exists when  
\begin{equation}
   \bm{t}_1 \cdot \bm{Z}_{0}^{\mathsf{T}} \bm{K} \hat{\bm{q}}(\bm{\varepsilon}) = 0, \quad
   \bm{t}_2 \cdot \bm{Z}_{0}^{\mathsf{T}} \bm{K} \hat{\bm{q}}(\bm{\varepsilon}) = 0,
\end{equation}
two conditions that are certainly verified, so that the linear problem \eqref{eq_homogenization_x} can be solved for $\tilde{\bm{q}}^{*}=\tilde{\bm{q}}^{*}(\bm{\varepsilon})$. 
The solution is therefore found from equation \eqref{perotta} in the form 
\begin{equation}
    \bm{q}(\bm{\varepsilon}, \tilde{\bm{q}}^{*}(\bm{\varepsilon})) = 
    \hat{\bm{q}}(\bm{\varepsilon}) + \bm{Z}_{0} \tilde{\bm{q}}^{*}(\bm{\varepsilon}) \, ,
\end{equation}
linear in $\bm{\varepsilon}$, so that the elastic strain energy stored in a unit cell of the grid is a quadratic function of the applied uniform strain $\bm{\varepsilon}$
\begin{equation}
\label{cazzottone}
    \mathcal{E}(\bm{\varepsilon}) = 
    \frac{1}{2}\bm{q}(\bm{\varepsilon}, \tilde{\bm{q}}^{*}(\bm{\varepsilon}))
    \cdot 
    \bm{K}\bm{q}(\bm{\varepsilon}, \tilde{\bm{q}}^{*}(\bm{\varepsilon})). 
\end{equation}

By assuming a \emph{linear elastic} constitutive law for the equivalent continuum, the stress tensor $\bm{\sigma}$ is linearly related to the strain tensor $\bm{\varepsilon}$ through the fourth-order elasticity tensor $\mathbb{E}$, i.e. $\bm{\sigma} = \mathbb{E}\, \bm{\varepsilon}$, and the strain energy $\mathcal{W}$ is defined as 
\begin{equation}
\label{eq_const_incremental}
    \mathcal{W}(\bm{\varepsilon}) = \frac{1}{2} \bm{\varepsilon} \cdot \mathbb{E}\, \bm{\varepsilon}\, .
    %~~~~
    % \text{where} ~~~~
    % \bm{\sigma}=\mathbb{E}\bm{\varepsilon} \, .
\end{equation}
The elastic continuum equivalent to the grid can be obtained by imposing a match between the energies $\mathcal{E}$, eq.~\eqref{cazzottone}, and $\mathcal{W}$, eq.~\eqref{eq_const_incremental}, 
\begin{equation}
\label{sfig}
    \bm{\varepsilon} \cdot \mathbb{E}\, \bm{\varepsilon} = 
    \frac{1}{|\mathcal{A}|} \bm{q}(\bm{\varepsilon}, \tilde{\bm{q}}^{*}(\bm{\varepsilon})) 
    \cdot 
    \bm{K}\bm{q}(\bm{\varepsilon}, \tilde{\bm{q}}^{*}(\bm{\varepsilon})),
\end{equation}
where $|\mathcal{A}|$ represents the area of the unit cell. Eq. (\ref{sfig}) directly leads to the  constitutive tensor for the effective Cauchy material equivalent to the lattice 
\begin{equation}
\label{vecchiaccio}
    \mathbb{E} = 
    \frac{1}{2 |\mathcal{A}|} \frac{\partial^{2}\, \bm{q}(\bm{\varepsilon}, \tilde{\bm{q}}^{*}(\bm{\varepsilon}))
    \cdot 
    \bm{K} \bm{q}(\bm{\varepsilon}, \tilde{\bm{q}}^{*}(\bm{\varepsilon}))}{\partial \bm{\varepsilon}^2} \, ,
\end{equation}
equipped with the minor symmetries enforced by the symmetry of $\bm{\varepsilon}$ and the major symmetry related to the symmetry of $\bm{K}$.

\section{Influence of rigid elements on homogenized elastic properties }
\label{Sec_properties}

\begin{figure}[htb!]
\centering
 \includegraphics[scale=1, align=c]{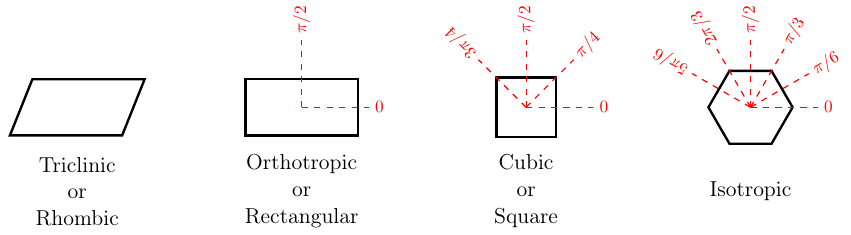}
\caption{Examples of microstructures  leading to equivalent 2D-continua enhanced with the symmetries shown in Table~\ref{tab_sym_group}. Reflection axes are indicated with red dashed lines.
}
\label{wlf}
\end{figure}

The purpose of this section is to show how the elasticity tensor is affected by the presence of rigid elements. The homogenization procedure illustrated in the previous section is applied to paradigmatic cases designed to cover all the material symmetry groups in a two-dimensional formulation. To this purpose, the following examples are inspired by Chinese lattices \cite{dye2012chinese}, which are composed of closed concentric elements within the unit cells. For each example the fourth-order elasticity tensor (\ref{vecchiaccio}) will be derived and expressed with respect to the reference system $\be_1$--$\be_2$. This tensor can also be written in the Voigt notation as a symmetric matrix 
\begin{equation}
    \begin{bmatrix}
    \sigma_{11} \\[10pt]
    \sigma_{22} \\[10pt]
    \sigma_{12}
    \end{bmatrix}
    =
    \begin{bmatrix}
    \mathbb{E}_{1111} & \mathbb{E}_{1122} & \mathbb{E}_{1112} \\[10pt]
    \cdot & \mathbb{E}_{2222} & \mathbb{E}_{2212} \\[10pt]
    \cdot & \cdot & \mathbb{E}_{1212}
    \end{bmatrix}
    \begin{bmatrix}
    \varepsilon_{11} \\[10pt]
    \varepsilon_{22} \\[10pt]
    2 \varepsilon_{12}
    \end{bmatrix}
    \notag \, .
\end{equation}
A \emph{symmetry transformation} for the equivalent solid is an orthogonal tensor $\bQ$ that leaves the components of the fourth-order tensor $\mathbb{E}$ unchanged, namely, 
\begin{equation}
\label{rotazione8}
    \mathbb{E}'_{ijkl} = Q_{pi} Q_{qj} Q_{rk} Q_{sl} \mathbb{E}_{pqrs} \equiv \mathbb{E}_{ijkl}, 
\end{equation}
where $\mathbb{E}'_{ijkl}$ are the components of $\mathbb{E}$ with respect to the rotated reference system $\be_i' = \bQ \be_i$.
The symmetry group for the equivalent solid is defined as the set of all orthogonal tensors $\bQ$ that obey \eqref{rotazione8}. In a two-dimensional context, where indices run between 1 and 2, only 4 symmetry groups exist \cite{forte1996symmetry,chadwick2001new}. Examples of structures are illustrated in Fig.~\ref{wlf}, forming elastic grids which lead to equivalent continua possessing the symmetries listed in Table \ref{tab_sym_group}.
%
%%%%%%%%%%%%%%%%%%%%%%%%%%%%%%%%%%%%%%%%%%%%%%%%%%%%%%%%%%%%%%%%%%%%
\begin{table}[htb!]
    \begin{center}  
    \begin{tabularx}{\textwidth}{@{} m{0.15\textwidth}<{\raggedright} m{0.7\textwidth}<{\centering}@{\hskip -0.4cm} m{0.15\textwidth}<{\centering} }
    \toprule
    Symmetry group & Restrictions on $\mathbb{E}$ & Independent elastic constants \\[5mm]
    \midrule \\
    Triclinic or Rhombic & None & 6 \\[7mm]
    Orthotropic or Rectangular & $\mathbb{E}_{1112}=\mathbb{E}_{2212}=0$ & 4  \\[7mm]
    Cubic or Square & $\mathbb{E}_{1112}=\mathbb{E}_{2212}=0 \, , \; \mathbb{E}_{1111}=\mathbb{E}_{2222}$ & 3  \\[7mm]
    Isotropic & $\mathbb{E}_{1112}=\mathbb{E}_{2212}=0 \, , \;  \mathbb{E}_{1111}=\mathbb{E}_{2222} \, , \; \mathbb{E}_{1212}=\displaystyle \frac{\mathbb{E}_{1111}- \mathbb{E}_{1122}}{2}$ & 2 \\ [3mm] 
    \bottomrule
    \end{tabularx}
    \caption{Material symmetry groups of the elasticity tensor $\mathbb{E}$ in a two-dimensional formulation.}
    \label{tab_sym_group}
    \end{center}
\end{table}
%%%%%%%%%%%%%%%%%%%%%%%%%%%%%%%%%%%%%%%%%%%%%%%%%%%%%%%%%%%%%%%%%%%%

%%%%%%%%%%%%%%%%%%%%%%%%%%%%%%%%%%%%%%%%%%%%%%%%%%%%%%%%%%%%%%%%%%%%
%%%%%%%%%%%%%%%%%%%%%%%%%%%%%%%%%%%%%%%%%%%%%%%%%%%%%%%%%%%%%%%%%%%%

For a triclinic material, the compliance matrix $\mathbb{S}=\mathbb{E}^{-1}$ reads as
\begin{equation}
    \mathbb{S}=
    \begin{bmatrix}
    \displaystyle \phantom{-} \frac{1}{E_{1}} & \displaystyle -\frac{\nu_{21}}{E_{2}} & \displaystyle \phantom{-} \frac{\eta_{1,12}}{G_{12}} \\[10pt]
    \displaystyle -\frac{\nu_{12}}{E_{1}} & \displaystyle \phantom{-} \frac{1}{E_{2}} & \displaystyle \phantom{-} \frac{\eta_{2,12}}{G_{12}} \\[10pt]
    \displaystyle \phantom{-} \frac{\eta_{12,1}}{E_{1}} & \displaystyle \phantom{-} \frac{\eta_{12,2}}{E_{2}} & \displaystyle \phantom{-} \frac{1}{G_{12}}
    \end{bmatrix}
    \notag \, , 
\end{equation}
namely, in terms of engineering constants as suggested by Lekhnitskii \cite{lekhnitskii1964theory}, in which $E_{i}$, $G_{ij}$, and $\nu_{ij}$ are respectively the Young's moduli, the shear moduli and Poisson's ratios measured in the $i$--$j$ directions; in addition, coefficients $\eta_{i,ij}=\varepsilon_{i}/\gamma_{ij}$ and $\eta_{ij,i}=\gamma_{ij}/\varepsilon_{i}$ denote the coefficients of mutual influence of I and II kind, respectively.

\subsection{Triclinic, orthotropic, and cubic materials} 
\begin{figure}[h!]
    \centering
    \begin{subfigure}[]{0.48\textwidth}
    \centering
    \includegraphics[width=\textwidth, align=c]{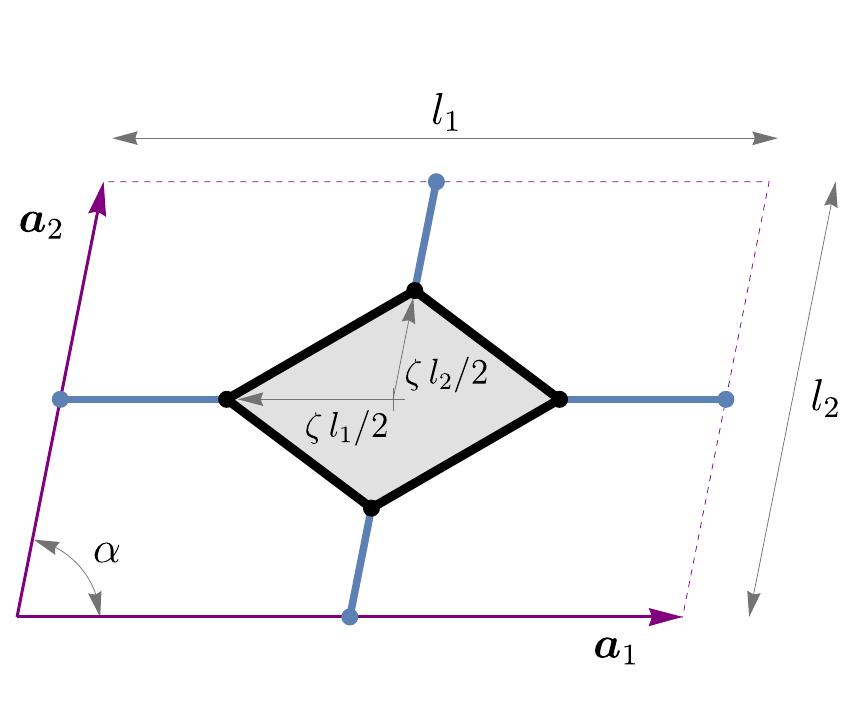}
    \end{subfigure}
    \begin{subfigure}[]{0.48\textwidth}
    \centering
    \includegraphics[width=\textwidth, align = c]{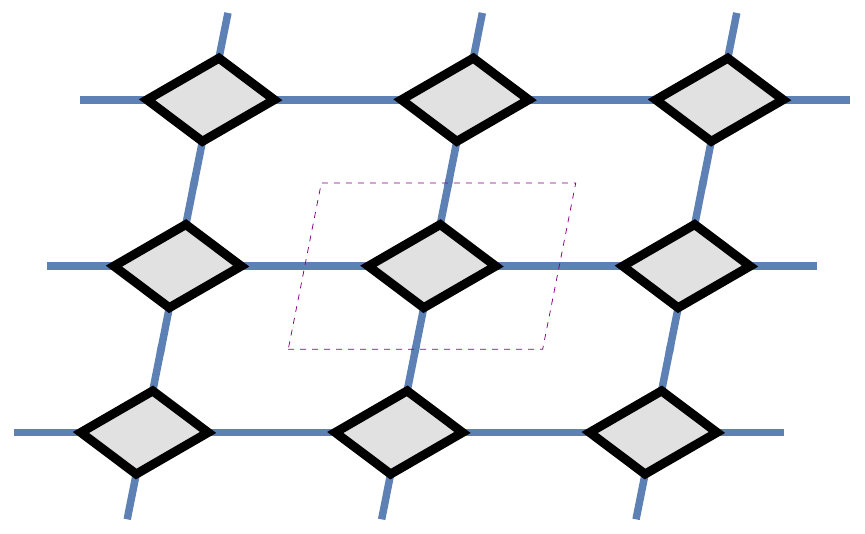}
    \end{subfigure}
    \caption{A unit cell composed of four elastic beams jointed through a parallelogrammic rigid inclusion (left) leads to a grid (right), which is equivalent to an elastic triclinic material. The latter reduces to an orthotropic elastic material when $\alpha=\pi/2$ and to a square material when $\alpha=\pi/2$ and $\xi=1$. The size of the rigid inclusion is set by the dimensionless parameter $\zeta \in [0,1]$.}
    \label{fig_triclinc_lattice}
\end{figure}

Triclinic, orthotropic, and cubic materials can be obtained from a lattice composed of the unit cell (with dimensions $ l_{1} \times l_{2}$) forming the elastic grid shown in Fig.~\ref{fig_triclinc_lattice}. 

Inside the unit cell, a parallelogrammic rigid inclusion is present, with its vertices rigidly jointed to four elastic beams. The components of the elasticity tensor, equivalent to the grid, have been made dimensionless through a division by the horizontal axial beam's stiffness 
\begin{equation}
    \tilde{\mathbb{E}}_{ijkl} = \mathbb{E}_{ijkl}/(EA/l_{1})\, \notag,  
\end{equation}
and depend on the extension of the rigid inclusion within the unit cell, via parameter $\zeta$, ranging between 0 (absence of rigid elements) and 1 (the whole unit cell becomes rigid). In particular, $\zeta$ represents the nondimensional length of the diagonals characterizing the parallelogrammic rigid inclusion, as sketched in Fig.~\ref{fig_triclinc_lattice}. Moreover, the following dimensionless ratios are used 
\begin{equation}
\label{eq_input_triclinic}
    \lambda_{1} = l_{1}(1-\zeta) \sqrt{\frac{A_{1}}{I_{1}}} \, , \quad 
    \lambda_{2}= l_{2}(1-\zeta) \sqrt{\frac{A_{2}}{I_{2}}} \, , \quad 
    \xi = l_{2}/l_{1} \,, \quad 
    \chi=A_{2}/A_{1} \, ,
\end{equation}
where the slenderness $\lambda_{1}$ and $\lambda_{2}$ depend on $\zeta$ because an increase in the dimension of the rigid inclusion leads to a reduction in the length of the elastic beams. In the following, the cross-section areas of the beams have been assumed to be equal, $A_{2}=A_{1}=A$, for simplicity. 
%
%%%%%%%%%%%%%%%%%%%%%%%%%%%%%%%%%%%%%%%%%%%%%%%%%%%%%%%%%%%%%%%%%%%%
%%%%%%%%%%%%%%%%%%%%%%%%%%%%%%%%%%%%%%%%%%%%%%%%%%%%%%%%%%%%%%%%%%%%

Except for $\alpha=\pi/2$, the homogenized constitutive tensor equivalent to the grid represents a triclinic material and is characterized by the following components (in the reference system $\be_1$--$\be_2$ shown in Fig. \ref{fig_unitcell_example}) 
{\allowdisplaybreaks
\begin{align}
\label{frittelle}
    \tilde{\mathbb{E}}_{1111} &= \frac{\sin\alpha}{1-\zeta} \biggl[ \frac{1+\xi}{\xi \sin^{2}\alpha}- 1 -\cos^2\alpha +\frac{12\cos^2\alpha}{\lambda_{2}^{2}+\lambda_{1}^{2}\xi} \biggr] \, , \quad
    \tilde{\mathbb{E}}_{1212} = \frac{\sin\alpha}{1-\zeta} \left[ \frac{12 \sin^{2}\alpha}{\lambda_{2}^{2}+\lambda_{1}^{2}\xi}+\cos^{2}\alpha \right] \, , \nonumber \\[10pt]
    \tilde{\mathbb{E}}_{1122} &= \frac{\cos^{2}\alpha \sin\alpha(\lambda_{2}^{2}+\lambda_{1}^{2} \xi-12)}{(1-\zeta)(\lambda_{2}^{2}+\lambda_{1}^{2} \xi)} \, , \quad 
    \tilde{\mathbb{E}}_{2212}  = \frac{\cos\alpha\sin^{2}\alpha(\lambda_{2}^{2}+\lambda_{1}^{2} \xi-12)}{(1-\zeta)(\lambda_{2}^{2}+\lambda_{1}^{2} \xi)}\, , \nonumber \\[10pt]
    \tilde{\mathbb{E}}_{1112} &= \frac{\cos^{3}\alpha [\lambda_{2}^{2}+\lambda_{1}^{2} \xi+12 \tan^{2}\alpha]}{(1-\zeta)(\lambda_{2}^{2}+\lambda_{1}^{2} \xi)} \, , \quad  
    \tilde{\mathbb{E}}_{2222}  = \frac{1}{1-\zeta}\left[ \sin^3\alpha + 3\frac{\sin\alpha +\sin(3\alpha)}{\lambda_{2}^{2}+\lambda_{1}^{2} \xi} \right] \, . 
% \tilde{\mathbb{E}}_{2222} &= \frac{3\sin\alpha (\lambda_{2}^{2}+\lambda_{1}^{2} \xi+4)-\sin(3 \alpha)(\lambda_{2}^{2}+\lambda_{1}^{2} \xi-12)}{4(1-\zeta)(\lambda_{2}^{2}+\lambda_{1}^{2} \xi)} \, .
\end{align}
}
In the particular case $\alpha=\pi/2$, the principal directions $\bm{a}_{1}$ and $\bm{a}_{2}$ of the cell are orthogonal, so that an orthotropic material is obtained
\begin{equation}
    \begin{bmatrix}
    \tilde{\mathbb{E}}_{1111} & \tilde{\mathbb{E}}_{1122} & \tilde{\mathbb{E}}_{1112} \\[10pt]
    \cdot & \tilde{\mathbb{E}}_{2222} & \tilde{\mathbb{E}}_{2212} \\[10pt]
    \cdot & \cdot & \tilde{\mathbb{E}}_{1212}
    \end{bmatrix}
    =
    \frac{1}{1-\zeta}
    \begin{bmatrix}
    \displaystyle \frac{1}{\xi} & 0 & 0 \\[10pt]
    \cdot &  \displaystyle 1 & 0 \\[10pt]
    \cdot & \cdot & \displaystyle \frac{12}{\lambda_{2}^{2} + \lambda_{1}^{2} \xi}
    \end{bmatrix} \, ,
\end{equation}
where the term $\tilde{\mathbb{E}}_{1122}=0$ is null because, when subjected to vertical (or horizontal) uniaxial stress, the horizontal (or vertical) deformation is null.  

Another particular case is obtained when $\alpha=\pi/2$ and $\xi=1$, which implies $\tilde{\mathbb{E}}_{1111} \equiv \tilde{\mathbb{E}}_{2222}$, so that a cubic material is obtained.
 
In terms of engineering constants (made dimensionless through division by $EA/l_{1}$), the components (\ref{frittelle}) provide the values 
{\allowdisplaybreaks
\begin{align}
\label{orchite}
    \tilde{E}_{1} &= \frac{\csc \alpha}{\xi(1-\zeta)} \, , \quad  \tilde{E}_{2} = \frac{24 \sin \alpha }{(1-\zeta)[ \lambda_{2}^{2}+\lambda_{1}^{2} \xi+12+\cos(2 \alpha)(\lambda_{2}^{2}+\lambda_{1}^{2} \xi-12) ]} \, , \nonumber \\[10pt]
    \tilde{G}_{12} &= \frac{12 \csc \alpha}{ (1-\zeta)[\lambda_{2}^{2}-12+(\lambda_{1}^{2}-12)\xi+12 \csc^{2}\alpha(1+\xi)]} \, , \nonumber \\[10pt]
    \eta_{1,12} &= \frac{12 \xi \cot \alpha}{12-\lambda_{2}^{2}+(12-\lambda_{1}^{2})\xi -12 \csc^{2}\alpha(1+\xi)} \, , \quad \eta_{2,12} =  \frac{\lambda_{2}^{2} +\lambda_{1}^{2}\xi -12 }{12 \xi} \eta_{1,12} \, , \nonumber \\[10pt]
    \eta_{12,1} &= \frac{\tilde{E}_{1}}{\tilde{G}_{12}} \eta_{1,12} \, , \quad \eta_{12,2} =\frac{\tilde{E}_{2}}{\tilde{G}_{12}} \eta_{2,12}\, .
\end{align}
}
A particularly interesting feature is that all coefficients of mutual influence are independent of the rigid phase, while they are only functions of the elastic properties of the beams and their inclination $\alpha$. Another interesting feature is that the limit of rigid material is obtained when $\zeta \longrightarrow 1$ with a 1st-order singularity.

\begin{figure}[h!]
\centering
 \includegraphics[scale=1, align=c]{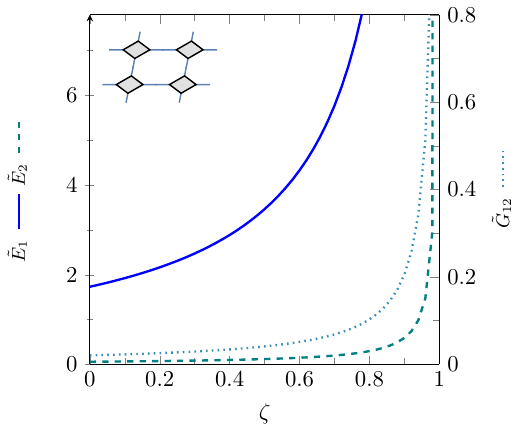}
\caption{
The elastic and shear moduli of a triclinic material grow and tend to infinity when the rigid elements invade the unit cell, $\zeta \rightarrow 1$. The unit cell is characterized by $\lambda_{1}=\lambda_{2}=20$, $\xi=2/3$, $\alpha=\pi/3$. Note the two different scales reporting $\tilde{E}_1$ and $\tilde{E}_2$ on the left and $\tilde{G}_{12}$ on the right of the figure.
}
\label{triclinic_stiffness_moduli_zeta}
\end{figure}

The effect of rigid inclusion is investigated in Fig.~\ref{triclinic_stiffness_moduli_zeta}, where the elastic and shear moduli, eqs.~(\ref{orchite})$_{1-3}$, are plotted versus $\zeta$. The figure pertains to the triclinic case, with $\lambda_{1}=\lambda_{2}=20$, $\xi=2/3$ and $\alpha=\pi/3$. As expected, all the constants blow up at the increase of $\zeta$, so that they approach infinity when $\zeta$ tends to 1.
%
%%%%%%%%%%%%%%%%%%%%%%%%%%%%%%%%%%%%
%%%%%%%%%%%%%%%%%%%%%%%%%%%%%%%%%%%%

\begin{figure}[h!]
%\centering
\begin{subfigure}[b]{0.45 \textwidth}
\includegraphics[scale=1, align =c]{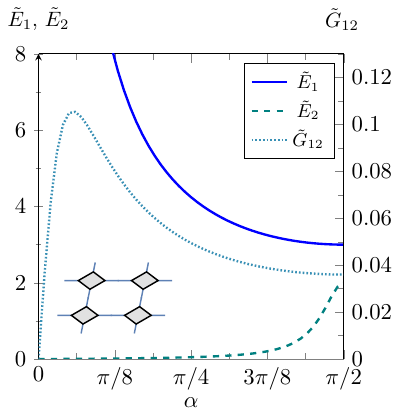}
\end{subfigure}
\hspace{0.2cm}
\begin{subfigure}[b]{0.45 \textwidth}
\includegraphics[scale=1, align=c]{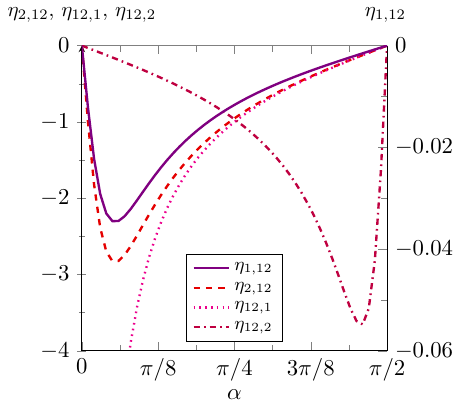}
\end{subfigure}
\caption{Engineering constants (\ref{orchite}) for an elastic material equivalent to the grid shown in Fig.~\ref{fig_triclinc_lattice}, as functions of the inclination of the elastic beams $\alpha$. The unit cell is characterized by $\lambda_{1}=\lambda_{2}=20$, $\xi=2/3$ and $\zeta=0.5$. Note the presence of two different scales reporting quantities on the left and on the right of the figures. }
\label{fig_plot_triclinic_eng_const_alpha}
\end{figure}

The effect of the inclination $\alpha$ of the elastic beams is investigated in Fig.~\ref{fig_plot_triclinic_eng_const_alpha} where $\lambda_{1}=\lambda_{2}=20$, $\xi=2/3$ and $\zeta=0.5$. 
The figure on the left shows that the elastic moduli $\tilde{E}_{1}$ and $\tilde{E}_{2}$ are affected by $\alpha$ in the opposite way, so that the former decreases, while the latter increases.
The shear modulus $\tilde{G}_{12}$ has a complicated behaviour, where after an increase a peak is reached, followed by a decreasing curve. 

The coefficients of mutual influence, reported on the right of Fig.~\ref{fig_plot_triclinic_eng_const_alpha}, do not depend on $\zeta$, thus the angle $\alpha$ represents the sole way to tune them.  
The coefficients of I kind, $\eta_{1,12}$ and $\eta_{2,12}$, show the same qualitative behaviour vanishing at the boundaries of the domain, while reaching a minimum for $\alpha \approx \pi/16$. In correspondence with this minimum, a shear deformation can maximize the consequent compression along the horizontal and vertical directions. The coefficients of the II kind behave differently from those pertinent to the I kind. Although $\eta_{12,2}$ assumes a shape analogous to that of coefficients of I kind, the minimum is shifted towards the right. The coefficient $\eta_{12,1}$ vanishes at $\alpha  = \pi/2$ and diverges towards the opposite side of the domain. Therefore, a tensile load along the vertical direction induces the maximum shear deformation when $ 3 \pi /8 < \alpha < \pi/2 $; otherwise, a tensile load along the horizontal direction leads to a shear deformation which grows when $\alpha$ is small.  
%
%%%%%%%%%%%%%%%%%%%%%%%%%%%%%%%%%%%%%%%
%%%%%%%%%%%%%%%%%%%%%%%%%%%%%%%%%%%%%%%%%%%

\subsection{Isotropic materials}
\begin{figure}[h!]
    \centering
    \begin{subfigure}[]{0.48\textwidth}
    \centering
    \includegraphics[width=\textwidth, align=c]{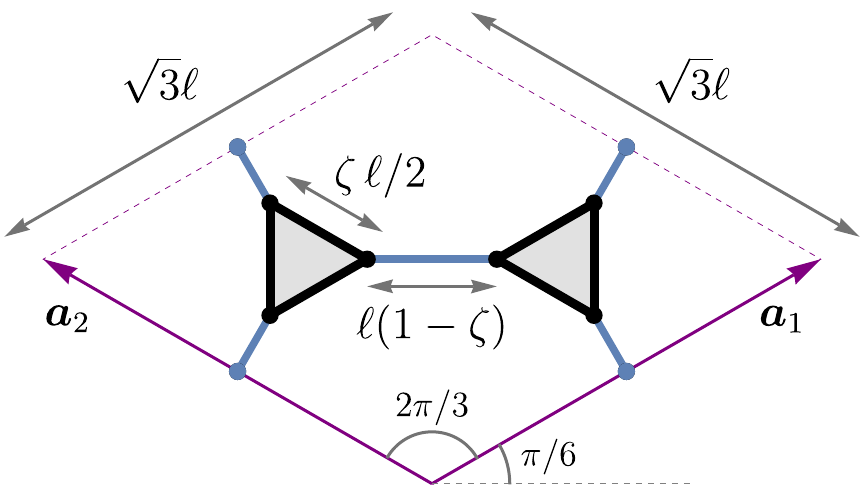}
    \end{subfigure}
    \begin{subfigure}[]{0.48\textwidth}
    \centering
    \includegraphics[width=\textwidth, align = c]{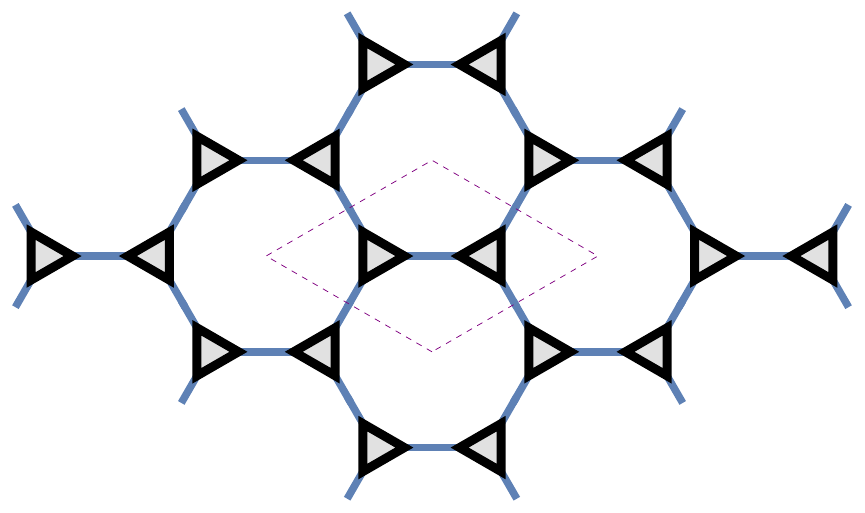}
    \end{subfigure}
    \caption{A unit cell containing two rigid elements in the form of equilateral triangles (left), with an edge of dimensionless measure $\zeta$, leads to an isotropic lattice (right).}
    \label{fig_hexagonal_lattice}
\end{figure}

\begin{figure}
    \centering
    \begin{subfigure}[]{0.48\textwidth}
    \centering
    \includegraphics[width=\textwidth, align=c]{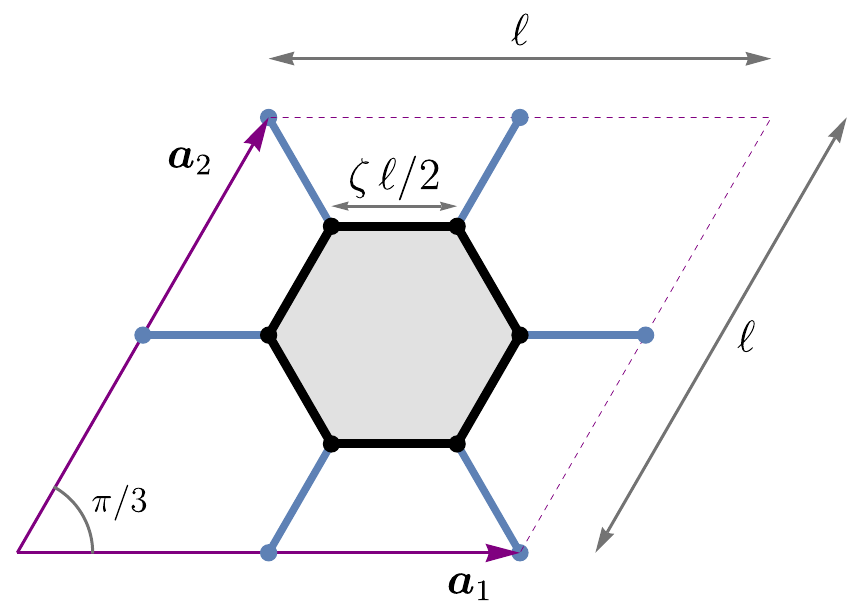}
    \end{subfigure}
    \begin{subfigure}[]{0.48\textwidth}
    \centering
    \includegraphics[width=\textwidth, align = c]{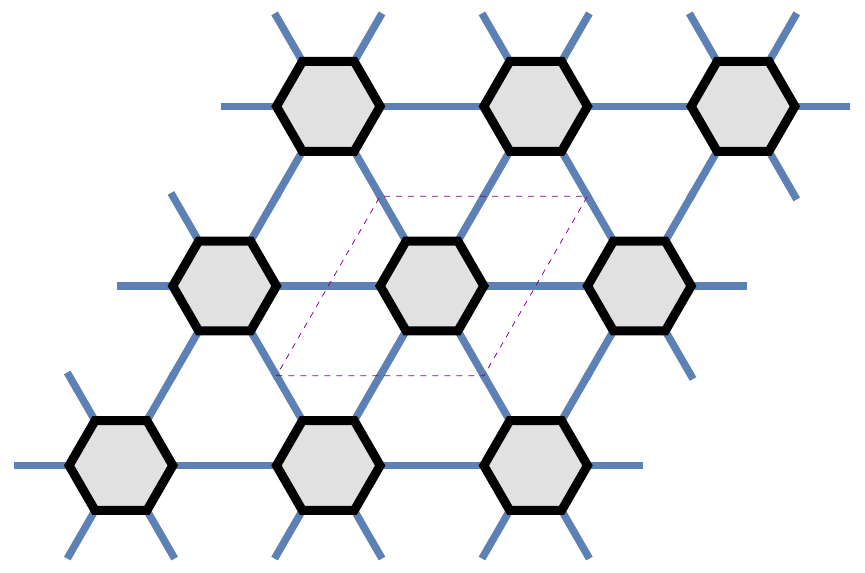}
    \end{subfigure}
    \caption{A unit cell containing a hexagonal rigid inclusion (left), with an edge of dimensionless measure $\zeta$, leads to an isotropic lattice (right).}
    \label{fig_triangular_lattice}
\end{figure}

Two different unit cells, both leading to an isotropic equivalent material, are investigated and compared, namely, a  cell containing two triangular rigid inclusions, Fig.~\ref{fig_hexagonal_lattice}, and another cell containing a hexagonal rigid inclusion, Fig.~\ref{fig_triangular_lattice}. In both cases, isotropy requires that the elastic beams (having cross-section area $A$, moment of inertia $I$,  and length $\ell$) are identical in the different directions, so that 
\begin{equation}
    \lambda= \ell(1-\zeta) \sqrt{\frac{A}{I}} \, ,
\end{equation}
where $\zeta \in [0,1]$ a dimensionless measure of the inclusion's edge, Fig.~\ref{fig_hexagonal_lattice}.
%
%%%%%%%%%%%%%%%%%%%%%%%%%%%%%%%%%%%%%%
%%%%%%%%%%%%%%%%%%%%%%%%%%%%%%%%%%%%%%

%%%%%%%%%%%%%%%%%%%%%%%%%%%%%%%%%%%%%%%
%%%%%%%%%%%%%%%%%%%%%%%%%%%%%%%%%%%%%

The elastic grid reported in Fig.~\ref{fig_hexagonal_lattice} (right) leads to an equivalent continuum characterized by the following components (divided by $EA/\ell$) of the elasticity tensor
\begin{equation}
\label{cavallo}
    \begin{bmatrix}
    \tilde{\mathbb{E}}_{1111} & \tilde{\mathbb{E}}_{1122} & \tilde{\mathbb{E}}_{1112} \\[10pt]
    \cdot & \tilde{\mathbb{E}}_{2222} & \tilde{\mathbb{E}}_{2212} \\[10pt]
    \cdot & \cdot & \tilde{\mathbb{E}}_{1212}
    \end{bmatrix}
    =
    \frac{1}{2 \sqrt{3} (1-\zeta)(\lambda^{2}+12)}
    \begin{bmatrix}
    \lambda^{2}+36  & \lambda^{2}-12 & 0 \\[10pt]
    \cdot & \lambda^{2}+36 & 0 \\[10pt]
    \cdot & \cdot & 24 
    \end{bmatrix} \, .
\end{equation}
Denoting with $\tilde{E}$ the dimensionless Young's modulus and with $\nu$ the Poisson's ratio, the components \eqref{cavallo} lead to 
\begin{equation}
    \tilde{E} = \frac{16 \sqrt{3}}{(1-\zeta)(\lambda^{2}+36)} \, , \quad \nu = 1-\frac{48}{\lambda^{2}+36} \, ,
\end{equation} 
showing that $\nu$ is independent of $\zeta$, but merely depends on $\lambda$. 

\begin{figure}[h!]
\centering
\begin{subfigure}[b]{0.47 \textwidth}
 \includegraphics[scale=1, align=c]{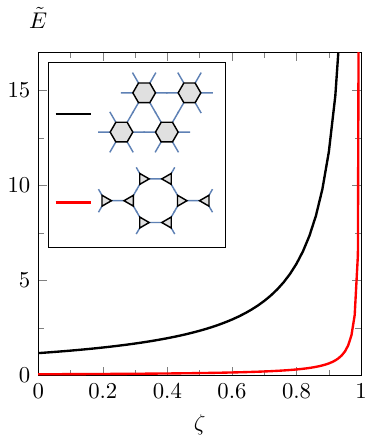}
\end{subfigure}
\begin{subfigure}[b]{0.47 \textwidth}
 \includegraphics[scale=1, align=c]{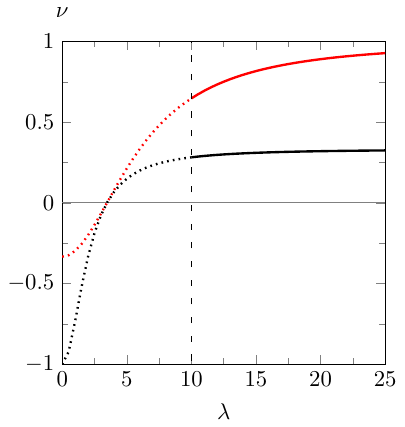}
\end{subfigure}
\caption{Young's modulus (as a function of $\zeta$) and Poisson's ratio (as a function of $\lambda$) for two elastic materials equivalent to the lattices with triangular (red line) and hexagonal (black line) inclusions.}
\label{fig_plot_eng_const}
\end{figure}

The lattice shown in Fig.~\ref{fig_triangular_lattice} (right), leads to the following components of the elasticity tensor
\begin{equation}
    \begin{bmatrix}
    \tilde{\mathbb{E}}_{1111} & \tilde{\mathbb{E}}_{1122} & \tilde{\mathbb{E}}_{1112} \\[10pt]
    \cdot & \tilde{\mathbb{E}}_{2222} & \tilde{\mathbb{E}}_{2212} \\[10pt]
    \cdot & \cdot & \tilde{\mathbb{E}}_{1212}
    \end{bmatrix}
    =
    \frac{\sqrt{3}}{4 \lambda^{2} (1 - \zeta)}
    \begin{bmatrix}
    3 (\lambda^{2}+4)  &  \lambda^{2}-12 & 0 \\[10pt]
    \cdot &  3 (\lambda^{2}+4)  & 0 \\[10pt]
    \cdot & \cdot & \lambda^{2}+12 
    \end{bmatrix} \, ,
\end{equation}
while the engineering constants are
\begin{equation}
    \tilde{E} = \frac{2(\lambda^{2}+12)}{\sqrt{3}(1-\zeta)(4+\lambda^{2})} \, , \quad \nu = \frac{1}{3}-\frac{16}{3(\lambda^{2}+4)} \, .
\end{equation}
The Young's modulus of the equivalent continuum is plotted in Fig.~\ref{fig_plot_eng_const} (on the left) as a function of $\zeta$, with $\lambda=20$; while the Poisson's ratio, independent of $\zeta$, is plotted as a function of $\lambda$ (on the right). 

In both cases, an auxetic behaviour, corresponding to negative Poisson's ratio, is found, with the following different limits for the two different cells
\begin{equation}
\label{tere}
    \begin{array}{ll}
    \displaystyle \nu \in \left(-\frac{1}{3}, \, 1\right)     &  \mbox{for the grid shown in Fig.~\ref{fig_hexagonal_lattice}}, \\ [3 mm]
    \displaystyle \nu \in \left(-1, \, \frac{1}{3}\right)     &  \mbox{for the grid shown in Fig.~\ref{fig_triangular_lattice}}. \\ 
    \end{array}
\end{equation}
It is important to notice that the above-introduced elastic modulus $E$ and Poisson's ratio $\nu$ refer to a \emph{purely two-dimensional theory}, so that they differ from their three-dimensional counterparts. In particular, the range of validity for  the constants $E$ and $\nu$ to enforce  positive definiteness of the strain energy is
\begin{equation}
    E>0 \, , ~~~ \nu \in (-1,\,1) \, .
\end{equation}
While the limits (\ref{tere}) apparently show that, with the two lattices, it is possible to cover all the range of $\nu$ for the equivalent continuum, it should be noticed that auxeticity is obtained for $lambda < 5$. However, for $\lambda < 10$, the beam theory cannot be considered valid and, as a consequence, the auxetic behaviour found here is not of practical interest.   
%
%%%%%%%%%%%%%%%%%%%%%%%%%%%%%%%
%%%%%%%%%%%%%%%%%%%%%%%%%%%%%%%%%
%

\subsection{Chiral lattices} 
\begin{figure}[h!]
    \centering
    \begin{subfigure}[]{0.48\textwidth}
    \centering
    \includegraphics[width=\textwidth, align=c]{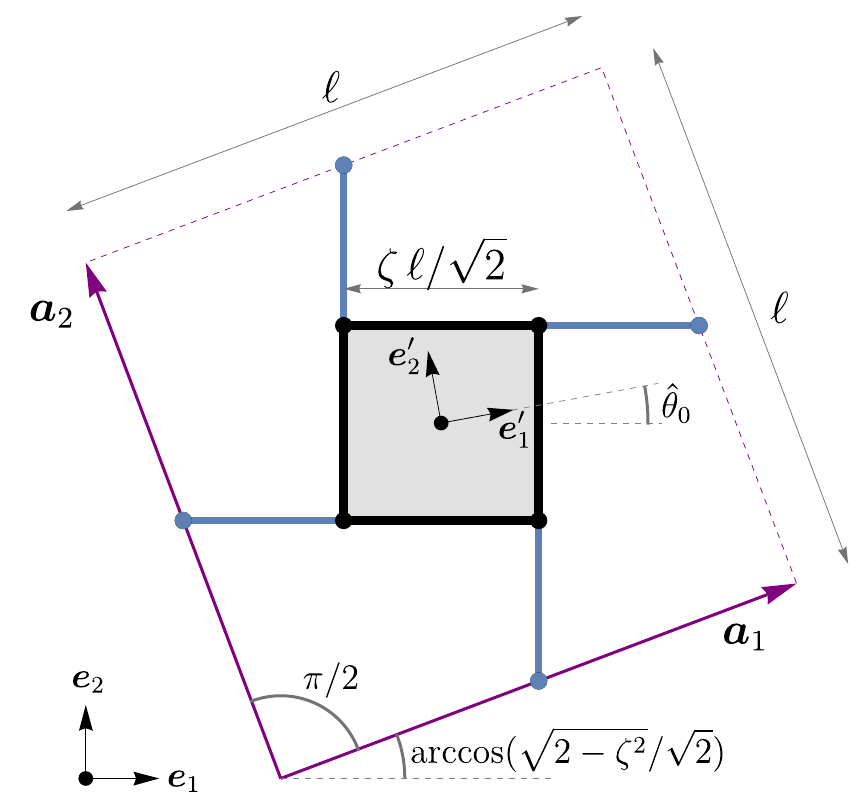}
    \end{subfigure}
    \begin{subfigure}[]{0.48\textwidth}
    \centering
    \includegraphics[width=\textwidth, align = c]{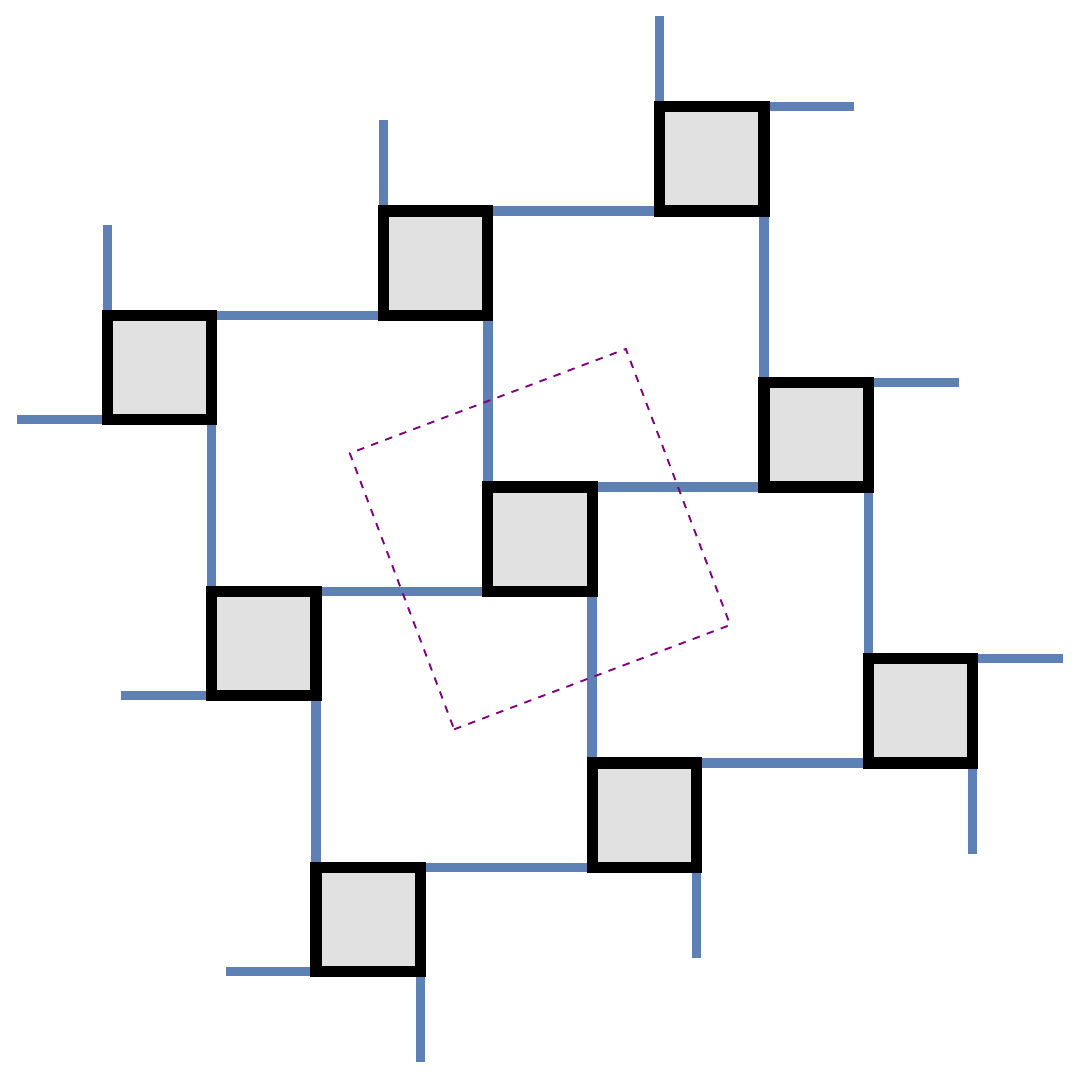}
    \end{subfigure}
    \caption{A chiral lattice is obtained from a square unit cell (with an edge of length $\ell$), containing a square rigid inclusion (with an edge of length $\zeta \ell/\sqrt{2}$, $\zeta \in [0,1]$). }
    \label{fig_square_chinese_askew_lattice}
\end{figure}

\begin{figure}
    \centering
    \begin{subfigure}[]{0.48\textwidth}
    \centering
    \includegraphics[width=\textwidth, align=c]{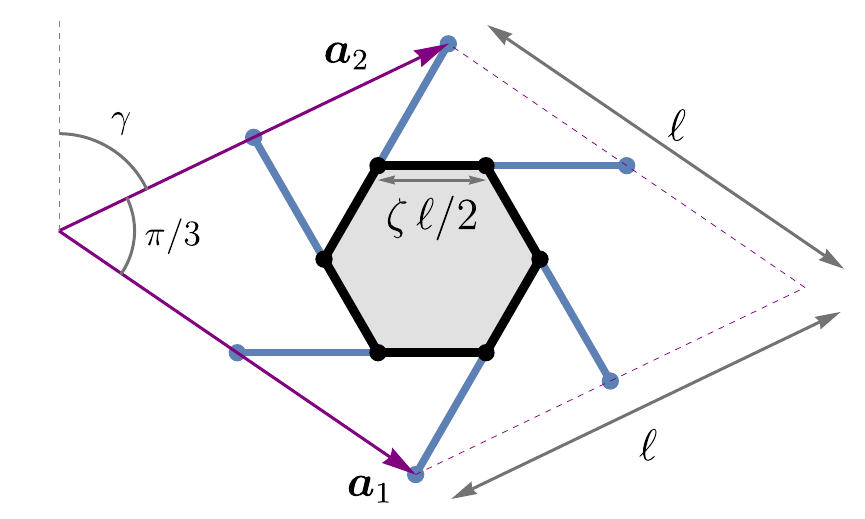}
    \end{subfigure}
    \begin{subfigure}[]{0.48\textwidth}
    \centering
    \includegraphics[width=\textwidth, align = c]{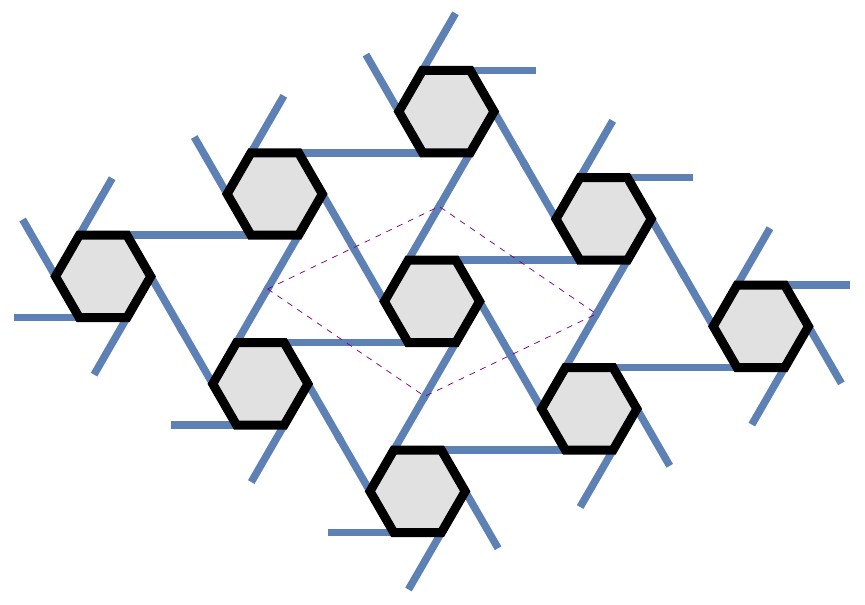}
    \end{subfigure}
    \caption{A chiral lattice is obtained from a rhombus unit cell (with an edge of length $\ell$), containing a hexagonal rigid inclusion with an edge of length $\zeta \ell/2$, $\zeta \in [0,1]$. The angle between vectors $\bm{a}_{1}$ and $\bm{a}_{2}$ is $\pi/3$ independently of the inclusion size $\zeta$. The angle between the vertical direction, $\be_2$, and $\ba_2$ is $\gamma=\arccos (\sqrt{3} \zeta / 2)$.}
    \label{fig_hexagon_chinese_askew_lattice}
\end{figure}

Two chiral lattices, inspired by the geometries reported in \cite{dye2012chinese}, are shown in Figs.~\ref{fig_square_chinese_askew_lattice} and \ref{fig_hexagon_chinese_askew_lattice}. They are homogenized by assuming the rigid elements as perfectly connected to the elastic beams. 
%
%%%%%%%%%%%%%%%%%%%%%%%%%%%%%%%%%
%%%%%%%%%%%%%%%%%%%%%%%%%%%%%%%%%

The former geometry involves a square unit cell, of dimensions $ \ell \times \ell$, Fig.~\ref{fig_square_chinese_askew_lattice}. Inside the unit cell, a rigid square element is defined through the length of its edge, $\zeta \ell/\sqrt{2}$, with $\zeta \in [0,1]$. Elastic beams, of slenderness 
\begin{equation}
    \lambda= \frac{\ell}{\sqrt{2}} \left( \sqrt{2-\zeta^{2}}-\zeta \right) \sqrt{\frac{A}{I}} \, ,
\end{equation}
are connected to the four corners of the inclusion. 

\begin{figure}[h!]
    \centering
    \begin{subfigure}[]{0.48\textwidth}
    \centering
    \includegraphics[width=\textwidth, align=c]{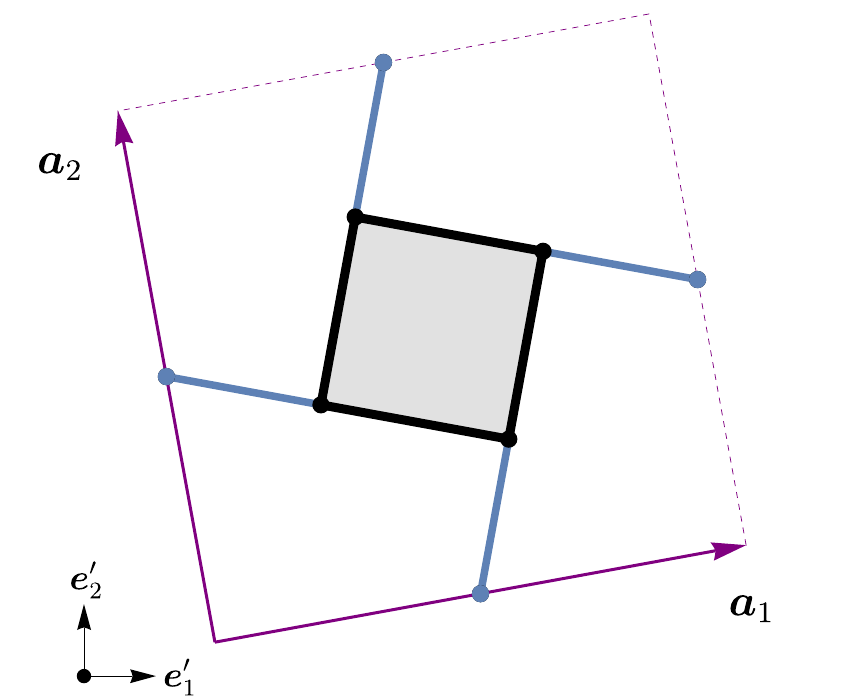}
    \end{subfigure}
    \begin{subfigure}[]{0.48\textwidth}
    \centering
    \includegraphics[width=\textwidth, align = c]{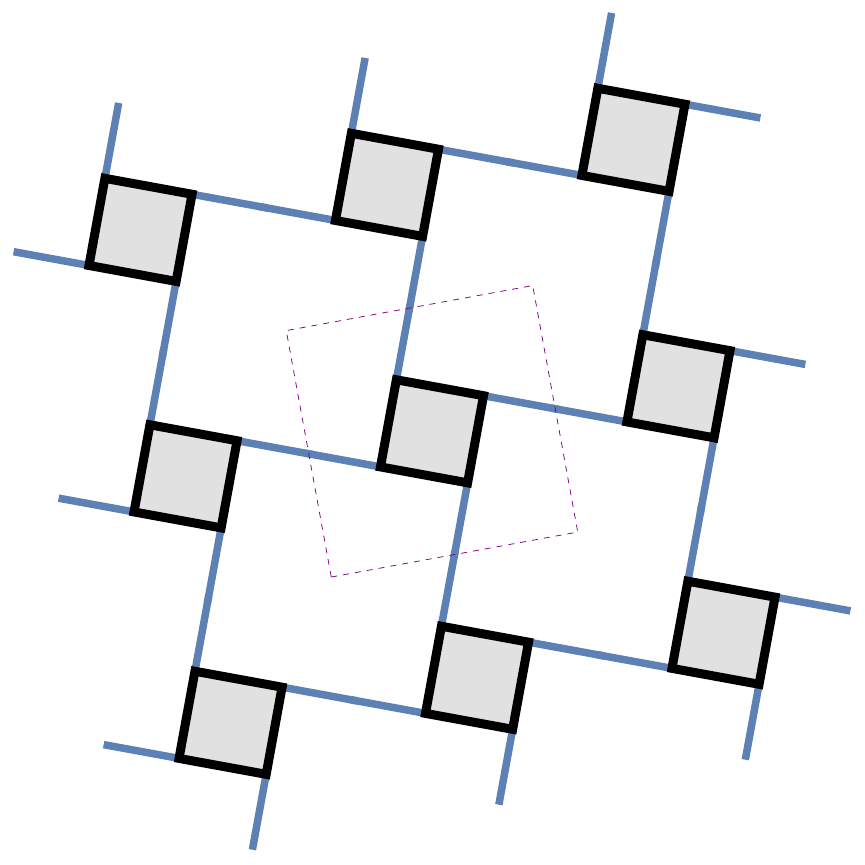}
    \end{subfigure}
    \caption{The unit cell corresponding to one of the principal reference systems, in which the material shown in Fig.~\ref{fig_square_chinese_askew_lattice} appears to be cubic, eqs. \eqref{eq_square_askew_chinese_lattice_principal}.}
    \label{fig_square_chinese_askew_lattice_principal}
\end{figure}

The homogenization leads to a material characterized by the following components of the elasticity tensor (divided by $EA/ \ell$), in the reference system $\be_1$--$\be_2$ shown in Fig.~\ref{fig_square_chinese_askew_lattice}, 
%
%{\allowdisplaybreaks
\begin{align}
\label{eq_square_askew_chinese_lattice}
    \tilde{\mathbb{E}}_{1111} &=\tilde{\mathbb{E}}_{2222} = \frac{(\zeta+\sqrt{2-\zeta^{2}})[96 \lambda^{2}-\zeta^{2}(\zeta^{2}-2)(\lambda^{2}-12)^{2}]}{4 \sqrt{2} \lambda^{2} (1-\zeta^{2}) \left[24+\zeta^{2}(\lambda^{2}-12) \right]}\, , \nonumber \\[10pt]
    \tilde{\mathbb{E}}_{1212} &= \frac{\left( \zeta +\sqrt{2-\zeta^{2}} \right) \left[24+\zeta^{2}(\lambda^{2}-12) \right]}{4 \sqrt{2} \lambda^{2}(1-\zeta^{2})} \, , \quad \tilde{\mathbb{E}}_{1122} = \frac{\zeta^{2}(\zeta^{2}-2) (\lambda^{2}-12)^{2}}{ [24+\zeta^{2}(\lambda^{2}-12)]^{2}} \tilde{\mathbb{E}}_{1212} \, , \nonumber \\[10pt]
    \tilde{\mathbb{E}}_{1112} &=\frac{\zeta \sqrt{2-\zeta^{2}} (\lambda^{2}-12)}{24+\zeta^{2}(\lambda^{2}-12)}\tilde{\mathbb{E}}_{1212} \, , \quad \tilde{\mathbb{E}}_{2212} =-\tilde{\mathbb{E}}_{1112} \, . 
\end{align}
%}
%
A rotation of the reference system $\be_1$--$\be_2$ by one of the anticlockwise angles
\begin{equation}
   \hat \theta_n = \arctan \frac{\sqrt{2} - \sqrt{2 - \zeta^2}}{\zeta} \pm n\frac{\pi}{4}, \quad \forall n \in \mathbb{Z},
\end{equation}
provides a principal reference system $\be_1'$--$\be_2'$ in which the components of the elasticity tensor show that the material is cubic
\begin{align}
\label{eq_square_askew_chinese_lattice_principal}
    \tilde{\mathbb{E}}'_{1111} &= \tilde{\mathbb{E}}'_{2222} = \frac{(\zeta+\sqrt{2-\zeta^{2}})[48 + \zeta^{2}(\lambda^{2}-12)]}{2 \sqrt{2} (1-\zeta^{2}) \left[24+\zeta^{2}(\lambda^{2}-12) \right]}\, , \nonumber \\[10pt]
    \tilde{\mathbb{E}}'_{1212} &= \frac{ 3\sqrt{2} \left( \zeta + \sqrt{2-\zeta^{2}} \right)}{\lambda^{2}(1-\zeta^{2})} \, , \quad \tilde{\mathbb{E}}'_{1122} = -\frac{\zeta^{2} \left(\zeta + \sqrt{2 - \zeta^{2}}\right) (\lambda^{2}-12)}{ 2 \sqrt{2} (1 - \zeta^2) [24 + \zeta^{2}(\lambda^{2}-12)]} \, , \nonumber \\[10pt]
    \tilde{\mathbb{E}}'_{1112} &= \tilde{\mathbb{E}}'_{2212} = 0 \, . 
\end{align}
In the principal reference system $\be_1'$--$\be_2'$ given by an anticlockwise rotation $\hat{\theta}_0$, the unit cell appears as shown in Fig.~\ref{fig_square_chinese_askew_lattice_principal}.
%
%%%%%%%%%%%%%%%%%%%%%%%%%%%%%%%%%%%%%%%%%%%
%%%%%%%%%%%%%%%%%%%%%%%%%%%%%%%%%%%%%%%%%%%%%%

\begin{figure}[h!]
\centering
\begin{subfigure}[b]{0.47 \textwidth}
 \includegraphics[scale=1, align=c]{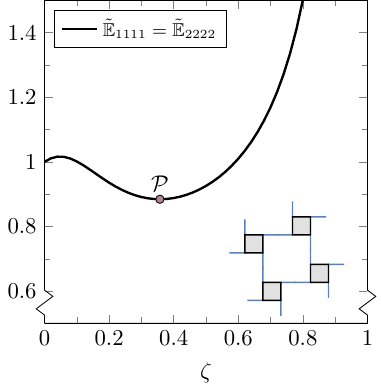}
\end{subfigure}
\hfill
\begin{subfigure}[b]{0.47 \textwidth}
 \includegraphics[scale=1, align=c]{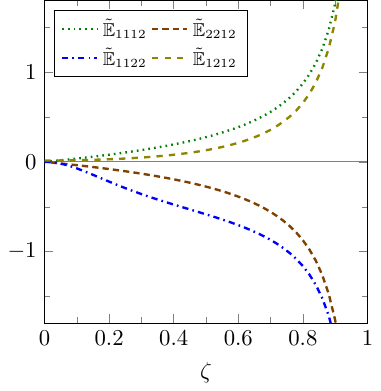}
\end{subfigure}
\caption{Components of the elasticity tensor, eq. (\ref{eq_square_askew_chinese_lattice}), for the chiral grid shown in Fig. \ref{fig_square_chinese_askew_lattice} as functions of the size of the square rigid inclusion $\zeta \in [0,1]$.  }
\label{fig_components_zeta_square_askew_chinese}
\end{figure}

\begin{enumerate}
    \item The dependence of the components \eqref{eq_square_askew_chinese_lattice} on the size $\zeta$ of the rigid inclusion can be better analysed in Fig.~\ref{fig_components_zeta_square_askew_chinese}, where $\lambda=20$.
\end{enumerate}

At increasing $\zeta$, the components $\tilde{\mathbb{E}}_{1111}= \tilde{\mathbb{E}}_{2222}$ exhibit an initial local maximum, followed by a minimum $\mathcal{P}$ (located at $\zeta \approx 0.36$) and subsequently blow-up. This behaviour is the result of different effects: when the rigid inclusion is very small, the elastic beams are practically aligned parallel to each other so that they provide the maximum axial stiffness. At increasing $\zeta$, the not-negligible distance $\zeta \ell/\sqrt{2}$ between the elastic beams generates a moment acting on the square inclusion, which represents an element of compliance inside the cell, leading the homogenized stiffness to drop. Finally, when the dimension of the rigid elements tends to occupy all the area of the unit cell, the stiffness diverges to infinity. 

The other components, reported in Fig.~\ref{fig_components_zeta_square_askew_chinese} on the left, simply increase or decrease following $\zeta$. 
%
%%%%%%%%%%%%%%%%%%%%%%%%%%%%%%%%%%%
%%%%%%%%%%%%%%%%%%%%%%%%%%%%%%%%%

Note that the components  $\tilde{\mathbb{E}}_{2212}$ and $\tilde{\mathbb{E}}_{1112}$ present in eq.~\eqref{eq_square_askew_chinese_lattice}$_{5}$ are identical, except for their sign. They are related to the Lekhnitskii coefficients of mutual influence, which depend on the clockwise/counterclockwise direction of the chirality in the unit cell. All the engineering constants are reported in Table \ref{cane}.
%
%%%%%%%%%%%%%%%%%%%%%%%%%%%%%%%%%
\begin{table}[!h]
    \begin{center}  
    \begin{tabularx}{\textwidth}{@{}  m{0.2\textwidth}<{\raggedright}@{\hskip -0.8cm} m{0.6\textwidth}<{\centering}@{\hskip -0.6cm} m{0.3\textwidth}<{\centering} }
    \toprule
    Orientation of chirality & \includegraphics[width=2.cm, height=2.cm]{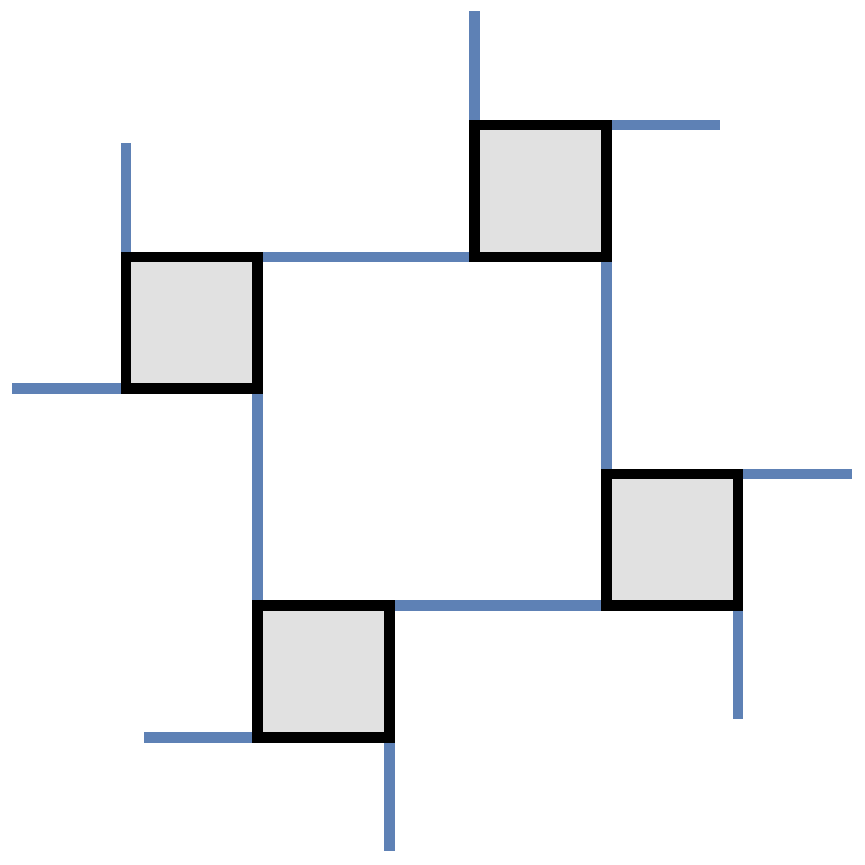} & \includegraphics[width=2.cm, height=2.cm]{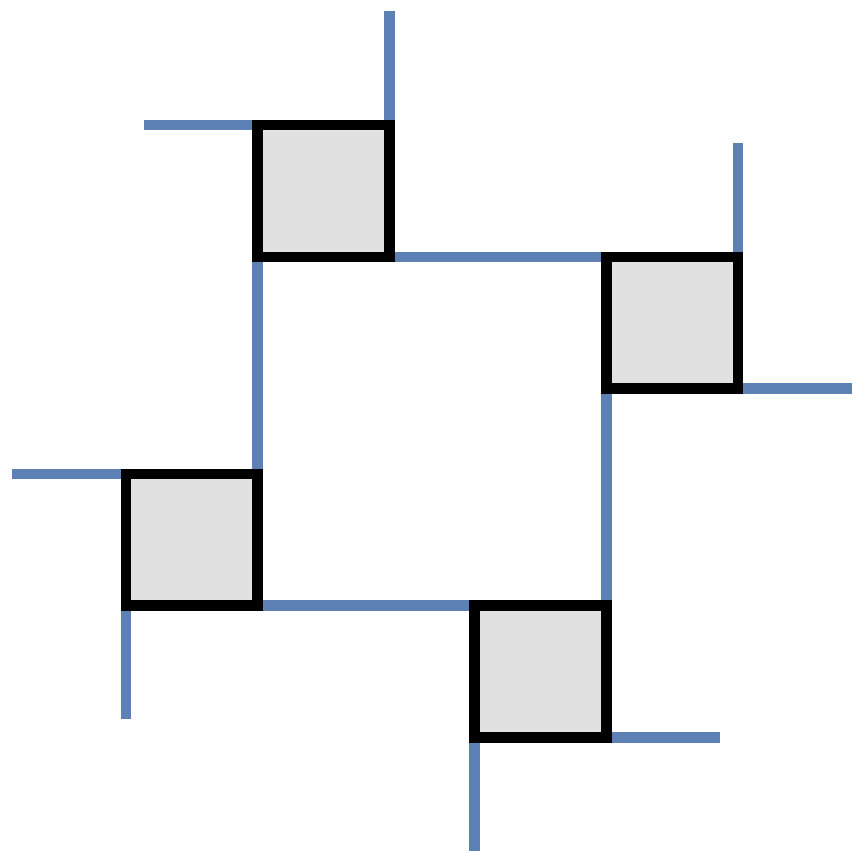} \\ [5mm]
    \midrule \\
    $\tilde{E}_{1}=\tilde{E}_{2}$ & $\displaystyle \frac{12 \sqrt{2} (\zeta+\sqrt{2-\zeta^{2}})}{(1-\zeta^{2})[24 +\zeta^{2}(\lambda^{2}-12)]}$ & $\displaystyle \frac{12 \sqrt{2} (\zeta+\sqrt{2-\zeta^{2}})}{(1-\zeta^{2})[24 +\zeta^{2}(\lambda^{2}-12)]}$  \\[8mm]
    $\tilde{G}_{12}$ & $\displaystyle \frac{6 \sqrt{2} (\zeta+\sqrt{2-\zeta^{2}})}{(1-\zeta^{2})[2 \lambda^{2} -\zeta^{2}(\lambda^{2}-12)]}$ & $\displaystyle \frac{6 \sqrt{2} (\zeta+\sqrt{2-\zeta^{2}})}{(1-\zeta^{2})[2 \lambda^{2} -\zeta^{2}(\lambda^{2}-12)]}$  \\[8mm]
    $\nu_{12}=\nu_{21}$ & $0$ & $0$  \\[8mm]
    $\eta_{1,12}=-\eta_{2,12}$ & $\displaystyle \frac{\zeta \sqrt{2-\zeta^{2}}(\lambda^{2}-12)}{2 \zeta^{2}(\lambda^{2}-12)-4 \lambda^{2}}$ & $\displaystyle - \frac{\zeta \sqrt{2-\zeta^{2}}(\lambda^{2}-12)}{2 \zeta^{2}(\lambda^{2}-12)-4 \lambda^{2}}$  \\[8mm]
    $\eta_{12,1}=-\eta_{12,2}$ & $\displaystyle - \frac{\zeta \sqrt{2-\zeta^{2}}(\lambda^{2}-12)}{\zeta^{2}(\lambda^{2}-12)+24}$ & $\displaystyle  \frac{\zeta \sqrt{2-\zeta^{2}}(\lambda^{2}-12)}{\zeta^{2}(\lambda^{2}-12)+24}$ \\ [3mm] 
    \bottomrule
    \end{tabularx}
    \caption{Comparison between the engineering constants for two square lattices with different (clockwise and counterclockwise) chirality.}
    \label{cane}
    \end{center}
\end{table}
%%%%%%%%%%%%%%%%%%%%%%%%%%%%%%%%%

Therefore, except for the coefficients of mutual influence, the chirality does not affect the characteristics of the equivalent elastic material. 

The last example of a chiral elastic grid inspired by a Chinese lattice is shown in Fig.~\ref{fig_hexagon_chinese_askew_lattice}. The unit cell, of rhombus shape, contains a rigid hexagonal element with edges of length $\zeta \ell$. 
%
%%%%%%%%%%%%%%%%%%%%%%%%%%%%%%%%%%%
%%%%%%%%%%%%%%%%%%%%%%%%%%%%%%%%%%%%%%%

\begin{figure}[h!]
\centering
\begin{subfigure}[b]{0.47 \textwidth}
\includegraphics[scale=1, align=c]{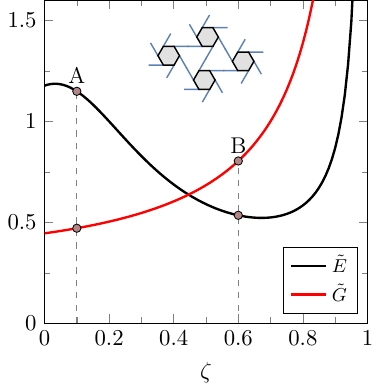}
\end{subfigure}
\begin{subfigure}[b]{0.47 \textwidth}
\includegraphics[scale=1, align=c]{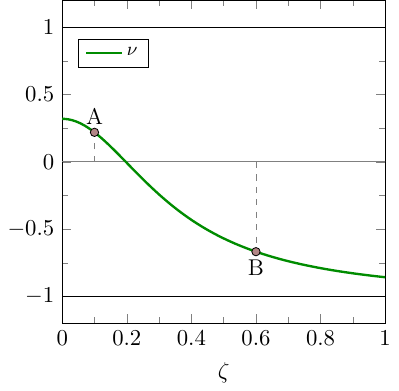}
\end{subfigure}
\caption{Young and shear moduli and Poisson's ratio, eq. \eqref{eq_eng_const_hexagonal_chiral}, for the chiral grid shown in Fig.~\ref{fig_hexagon_chinese_askew_lattice} as functions of the size of the hexagonal rigid inclusion $\zeta \in [0,1]$ with $\lambda=20$.}
\label{fig_engconstpoisson_hex_chinese}
\end{figure}

In this case, the slenderness of elastic beams is defined as
\begin{equation}
    \lambda=\frac{1}{2}\ell \left( \sqrt{4-3 \zeta^{2}} - \zeta \right) \sqrt{\frac{A}{I}} \, .
\end{equation}
The homogenization procedure returns an isotropic material, as chirality does not influence it (first-order). The equivalent material is characterized by the following non-vanishing components of the elasticity tensor (divided by  $EA/ \ell$)
{\allowdisplaybreaks
\begin{align}
    \tilde{\mathbb{E}}_{1111} &=\tilde{\mathbb{E}}_{2222}=\frac{\sqrt{3} \left( \zeta+\sqrt{4-3\zeta^{2}} \right)[48(\lambda^{2}+4)+\zeta^{2}(\lambda^{4}-144)]}{8\lambda^{2}(1-\zeta^{2})[\zeta^{2}(\lambda^{2}-12)+16]} \, , \nonumber \\[10pt]
    \tilde{\mathbb{E}}_{1122} &= \frac{\sqrt{3} \left( \zeta+\sqrt{4-3\zeta^{2}} \right)[\zeta^{2}(12+\lambda^{2})-16](12-\lambda^{2})}{8\lambda^{2}(1-\zeta^{2})[\zeta^{2}(\lambda^{2}-12)+16]} \, , \nonumber \\[10pt]
    \tilde{\mathbb{E}}_{1212}&= \frac{\sqrt{3} \left(\zeta +\sqrt{4-3 \zeta^{2}} \right) (12+\lambda^{2})}{8 \lambda^{2}(1-\zeta^{2})} \, . \label{eq_hexagon_askew_chinese_lattice}
    %\tilde{\mathbb{E}}_{1112}&=\tilde{\mathbb{E}}_{2212} = 0 \, .
\end{align}
}
Being the equivalent elastic material isotropic, the (dimensionless) engineering constants can be evaluated as
\begin{align}
\label{eq_eng_const_hexagonal_chiral}
    \tilde{E} &= \frac{16 \sqrt{3} \left( \zeta +\sqrt{4-3 \zeta^{2}} \right) (12+\lambda^{2})}{(1-\zeta^{2})[48(\lambda^{2}+4)+\zeta^{2}(\lambda^{4}-144)]} \, , \;
    \tilde{G} = \frac{\sqrt{3} \left( \zeta + \sqrt{4-3\zeta^{2}} \right)(12+\lambda^{2})}{8 \lambda^{2}(1-\zeta^{2})} \, , \nonumber \\[10pt]
    \nu &= \frac{64 \lambda^{2}}{48(\lambda^{2}+4)+\zeta^{2}(\lambda^{4}-144)}-1\, . 
\end{align}
The Young's modulus and the shear modulus are reported in Fig.~\ref{fig_engconstpoisson_hex_chinese} on the left as functions of $\zeta$ obtained with $\lambda=20$. Both of them diverge for $\zeta \rightarrow 1$ but, while the shear modulus is just a monotonously increasingly function of $\zeta$, the Young's modulus initially increases to reach a local maximum, subsequently decreases to reach a minimum at $\zeta \approx 0.671953$ and finally blows up. 
In the same way, the similar behaviour found for the chiral unit cell shown in Fig.~\ref{fig_components_zeta_square_askew_chinese} can be explained. 

The Poisson's ratio is reported in Fig.~\ref{fig_engconstpoisson_hex_chinese} on the right as a function of $\zeta$. At increasing $\zeta$, the Poisson's ratio starts from a positive value to become negative at $\zeta  > 0.2$, so that an auxetic behaviour is found. This is related to the fact that, at high values of $\zeta$, the rotation of the rigid hexagons leads to a marked transverse displacement.
%
%%%%%%%%%%%%%%%%%%%%%%%%%%%%%%%%%%%%%
%%%%%%%%%%%%%%%%%%%%%%%%%%%%%%%%%%%%%%%

\begin{figure}[h!]
    \centering
    \begin{subfigure}[]{0.49\textwidth}
    \centering
    \includegraphics[width= \textwidth, align=c]{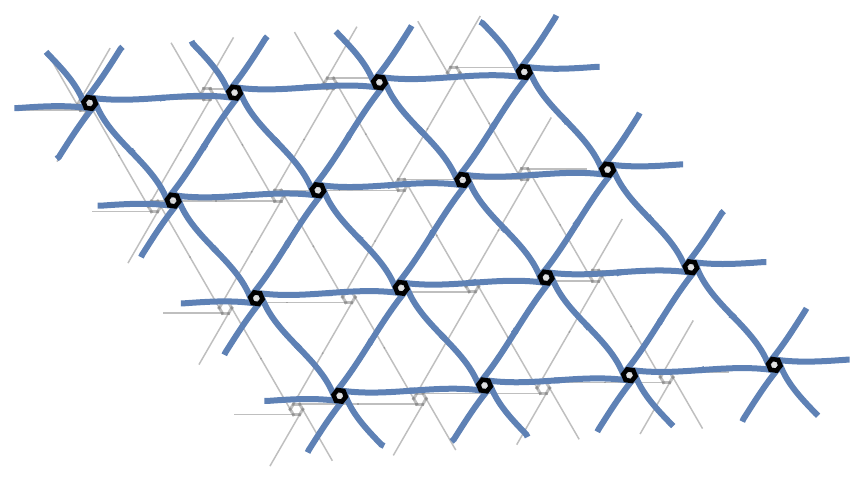}
    \end{subfigure}
    \hspace{-0.3cm}
    \begin{subfigure}[]{0.49\textwidth}
    \centering
    \includegraphics[width=\textwidth, align = c]{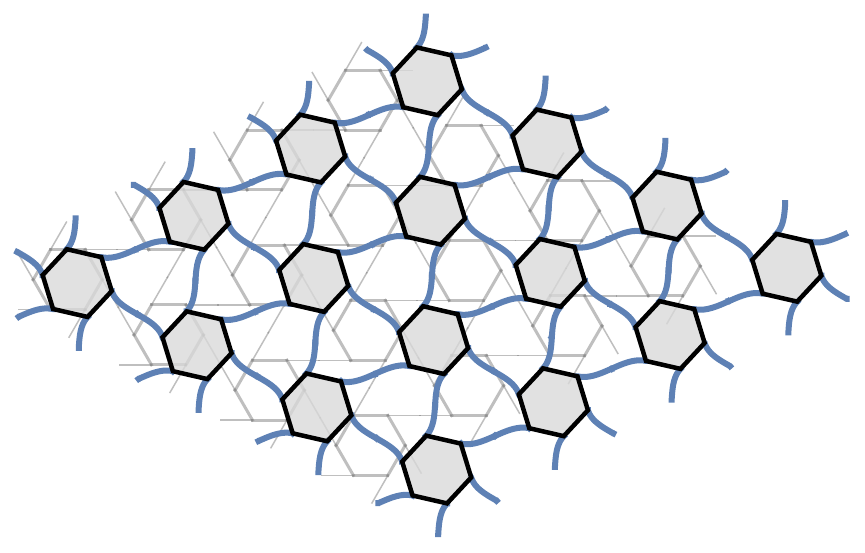}
    \end{subfigure}
    \caption{Deformed chiral lattice (with hexagonal rigid inclusions), subject to uniaxial stress aligned parallel to the horizontal direction. Small- ($\zeta=0.1$, upper part) and large- ($\zeta=0.6$, lower part) size inclusions. The undeformed configuration is reported as grey in the background.}
    \label{fig_lattice_hexagonal_chiral_deformed}
\end{figure}

For two different sizes of the hexagonal rigid inclusion (denoted as A, $\zeta=0.1$, and B, $\zeta=0.6$, in Fig.~\ref{fig_engconstpoisson_hex_chinese}), the respective deformed lattices, obtained through the imposition of an overall uniaxial stress aligned parallel to the horizontal direction, are reported in Fig.~\ref{fig_lattice_hexagonal_chiral_deformed}. Lower values of $\zeta$ allow for an axial stiffness higher than the shearing stiffness, because the elastic beams are almost aligned parallel to each other. At higher values of $\zeta$, the inclusions are subject to a not-negligible moment, so that they constitute an element of compliance, as the axial stiffness decreases and the homogenized material starts to show auxeticity.
%
%%%%%%%%%%%%%%%%%%%%%%%%%%%%%%%%%%%%%%%%%%%%%%
%%%%%%%%%%%%%%%%%%%%%%%%%%%%%%%%%%%%%%%%%%%%%%%%%%

\subsection{Influence of the connection between rigid and deformable elements}

\begin{figure}
    \centering
    \begin{subfigure}[]{0.48\textwidth}
    \centering
    \includegraphics[width=\textwidth, align=c]{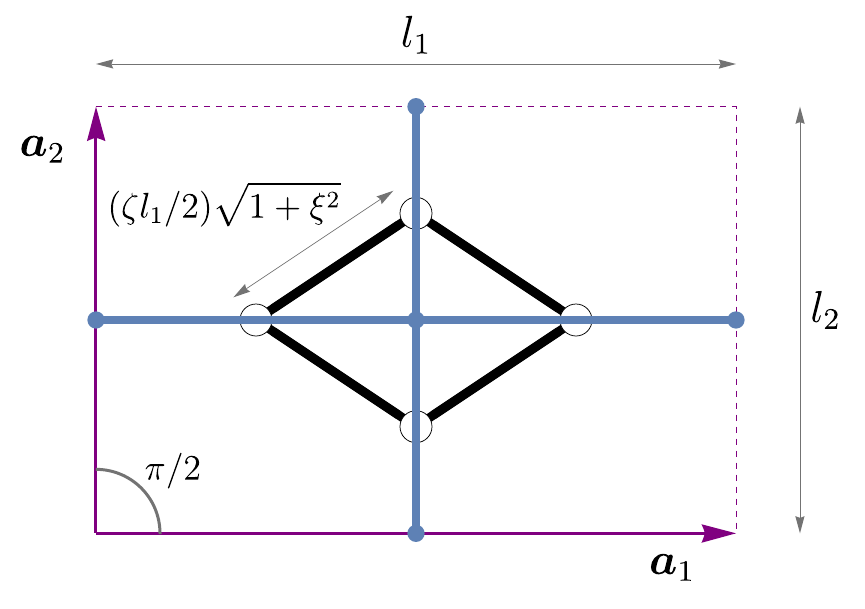}
    \end{subfigure}
    \begin{subfigure}[]{0.48\textwidth}
    \centering
    \includegraphics[width=\textwidth, align = c]{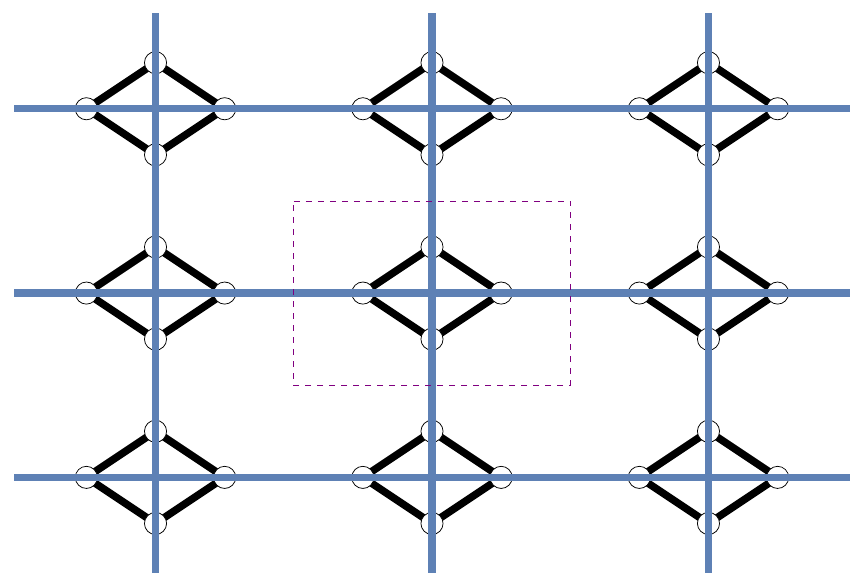}
    \end{subfigure}
    \caption{A lattice is obtained from a rectangular unit cell $l_{1}\times l_{2}$, containing four rigid rods of length $(\zeta l_{1}/2) \sqrt{1+\xi^{2}}$ with $\zeta \in [0,1]$.}
    \label{fig_rectangular_hinged}
\end{figure}

The effects of hinged connections between rigid and deformable elements are now investigated. To this purpose, the unit cell in Fig.~\ref{fig_triclinc_lattice} is reconsidered with $\alpha = \pi/2$, but now the parallelogrammic rigid inclusion is replaced by four distinct rigid rods of length $(\zeta l_{1}/2)\sqrt{1+\xi^{2}}$ (with $\xi=l_{1}/l_{2}$), which are \emph{hinged} to the elastic beams, as represented on the left of Fig.~\ref{fig_rectangular_hinged}. 
The area of the unit cell is $l_{1} \times l_{2}$, and the variation of $\zeta \in [0, \, 1]$ \lq shifts' the attaching points of the rigid rods along the elastic beams.
%
%%%%%%%%%%%%%%%%%%%%%%%%%%%%%%%%%%%%%%%%%%%%%%%%%%
%%%%%%%%%%%%%%%%%%%%%%%%%%%%%%%%%%%%%%%%%%%%%%%%%%
%
The dimensionless parameters defining such a unit cell read as
\begin{equation}
    \lambda_{1}= l_{1}(1-\zeta) \sqrt{\frac{A_{1}}{I_{1}}} \, , \quad \lambda_{2}= l_{2}(1-\zeta) \sqrt{\frac{A_{2}}{I_{2}}} \, , \quad \chi=A_{2}/A_{1} \, ,
\end{equation}
whereas the slendernesses $\lambda_{1}$ and $\lambda_{2}$ have been made to depend on $\zeta$ to facilitate comparison with the expressions \eqref{eq_input_triclinic}$_{1-2}$. For a straightforward interpretation of the following analytical results, the cross-sectional areas of the elastic beams are assumed to be equal, $A_{2}=A_{1}= A$. 

The hinged connection displays an interesting feature, namely, that the elastic rods comprised inside the hinged rhomboidal element can still deform. This feature will be shown to have interesting consequences. 

The homogenization procedure applied to the grid shown in Fig.~\ref{fig_rectangular_hinged} returns an elasticity tensor belonging to the rectangular symmetry group 
\begin{multline}
    \begin{bmatrix}
    \tilde{\mathbb{E}}_{1111} & \tilde{\mathbb{E}}_{1122} & \tilde{\mathbb{E}}_{1112} \\[10pt]
    \cdot & \tilde{\mathbb{E}}_{2222} & \tilde{\mathbb{E}}_{2212} \\[10pt]
    \cdot & \cdot & \tilde{\mathbb{E}}_{1212}
    \end{bmatrix}
    = \\
    \begin{bmatrix}
    \displaystyle \frac{\xi^{3}(1-\zeta)+1}{\xi(1+\xi^{3})(1-\zeta)} &  \displaystyle \frac{\xi \zeta}{(1+\xi^{3})(1-\zeta)} & 0 \\[10pt]
    \cdot & \displaystyle \frac{\xi^{3}+1-\zeta}{(1+\xi^{3})(1-\zeta)} & 0 \\[10pt]
    \cdot & \cdot & \displaystyle \frac{48}{(\lambda_{1}^{2} \xi+\lambda_{2}^{2})(4-\zeta)}
    \end{bmatrix} \, .
\end{multline}

The investigation is compared with the \lq clamped' version of the grid, shown in Fig.~\ref{fig_triclinc_lattice} ($\alpha=\pi/2$). The engineering constants relative to these two lattices are reported in Table~\ref{tab_eng_const_hinged_clamped}.
%
%%%%%%%%%%%%%%%%%%%%%%%%%%%%%%%%%%%%%%%%%%%%%%%%%%%%%
\begin{table}[!h]
    \begin{center}  
    \begin{tabularx}{\textwidth}{  m{0.15\textwidth}<{\centering} m{0.4\textwidth}<{\centering} m{0.4\textwidth}<{\centering}  }
    \toprule
    Unit cell & \includegraphics[width=3.cm]{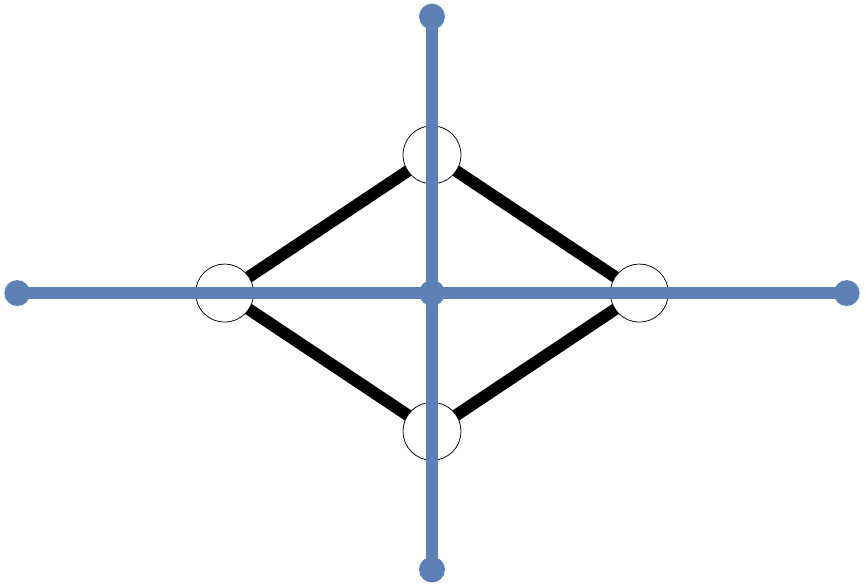} & \includegraphics[width=3.cm]{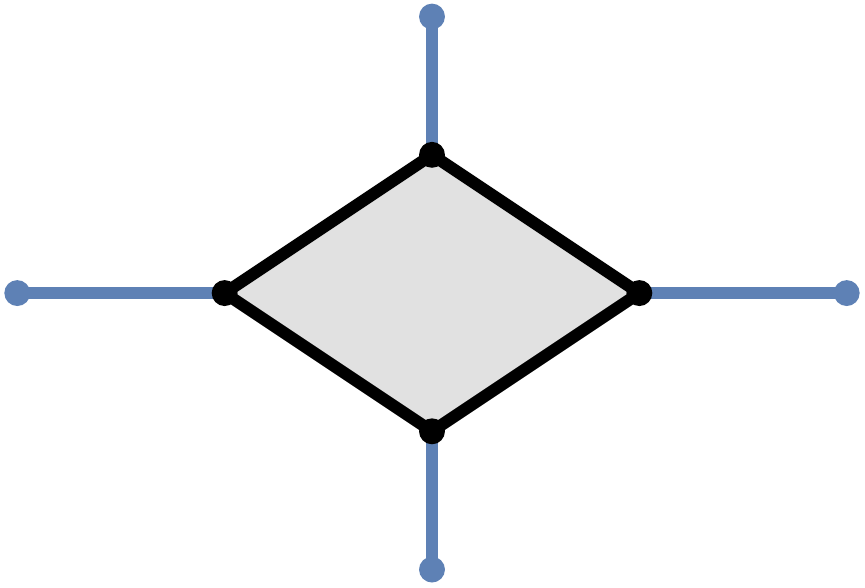} \\ [5mm]
    \midrule \\
    $\tilde{E}_{1}$ & $\displaystyle \frac{1+\xi^{3}}{\xi(\xi^{3}+1-\zeta)}$ & $\displaystyle \frac{1}{\xi(1-\zeta)}$  \\[8mm]
    $\tilde{E}_{2}$ & $\displaystyle \frac{1+\xi^{3}}{\xi^{3}(1-\zeta) +1}$ & $\displaystyle \frac{1}{1-\zeta}$  \\[8mm]
    $\tilde{G}_{12}$ & \scalebox{1}{ $\displaystyle \frac{48}{(\lambda_{1}^{2} \xi+\lambda_{2}^{2})(4-\zeta)} $} & \scalebox{1}{ $\displaystyle \frac{12}{(\lambda_{1}^{2} \xi+\lambda_{2}^{2})(1-\zeta)} $}  \\[8mm]
    $\nu_{12}$ & $\displaystyle \frac{\zeta \xi}{\xi^{3}+1-\zeta} $ & $0$  \\[8mm]
    $\nu_{21}$ & $\displaystyle \frac{\zeta \xi^{2}}{\xi^{3}(1-\zeta)+1} $  & $0$ \\[8mm]
    \bottomrule
    \end{tabularx}
    \caption{Engineering constants referred to the same rectangular unit cell containing four rigid rods. The latter are constrained to the elastic beams through \emph{hinges} (left, as in Fig.~\ref{fig_rectangular_hinged}) and through \emph{clamps} (right, as in Fig.~\ref{fig_triclinc_lattice}).}
    \label{tab_eng_const_hinged_clamped}
    \end{center}
\end{table}
%%%%%%%%%%%%%%%%%%%%%%%%%%%%%%%

\begin{figure}[h!]
\centering
\begin{subfigure}[b]{0.47 \textwidth}
\includegraphics[scale=1, align=c]{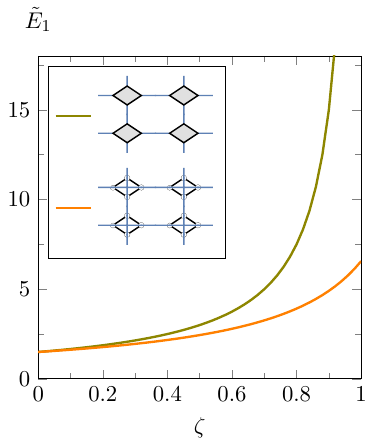}
\caption{}
\label{fig_plot_young_hinged_clamped}
\end{subfigure}
\begin{subfigure}[b]{0.47 \textwidth}
\includegraphics[scale=1, align=c]{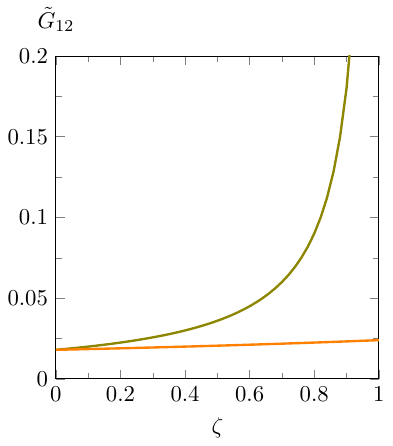}
\caption{}
\label{fig_plot_shear_hinged_clamped}
\end{subfigure}
\begin{subfigure}[b]{0.47 \textwidth}
\includegraphics[scale=1, align=c]{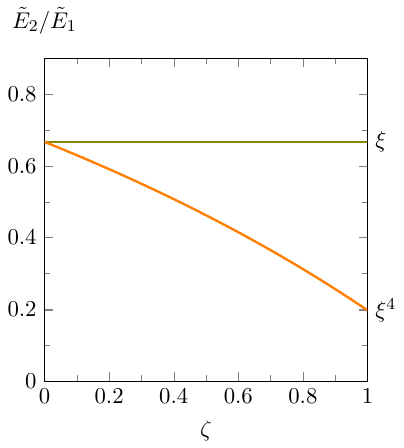}
\caption{}
\label{fig_plot_ratioyoung_hinged_clamped}
\end{subfigure}
\begin{subfigure}[b]{0.47 \textwidth}
\includegraphics[scale=1, align=c]{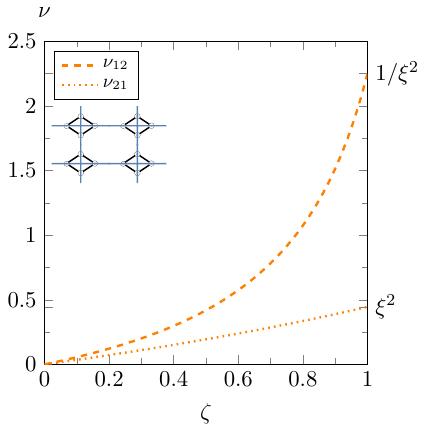}
\caption{}
\label{fig_plot_poisson_hinged}
\end{subfigure}
\caption{Engineering constants for the two rectangular lattices, with  \emph{clamped} and \emph{hinged} connections, plotted as functions of parameter $\zeta$ with $\lambda_{1}=\lambda_{2}=20$ and $\xi=2/3$.}
\label{fig_plot_eng_const_hinged}
\end{figure}

The Young $\tilde{E}_{1}$ and shear $\tilde{G}_{12}$ moduli are plotted in the upper part of Fig.~\ref{fig_plot_eng_const_hinged} as functions of $\zeta$ with $\lambda_{1}=\lambda_{2}=20$ and $\xi=2/3$. The values at $\zeta=0$ (denoting the absence of rigid inclusions) coincide for the two lattices. Both of them become stiffer at increasing $\zeta$: the moduli $\tilde{E}_{1}$ and $\tilde{G}_{12}$ of the clamped case grow rapidly and diverge when $\zeta \rightarrow 1$, because the unit cell becomes totally constrained. The moduli in the hinged case always assume finite values, because the lattice provides an elastic response even in the limit $\zeta\rightarrow 1$. As expected, the magnitude of axial as well as shearing stiffness is higher in the clamped case because of the stronger constraint. The ratio $\tilde{E}_{2}/\tilde{E}_{1}$ is plotted in Fig.~\ref{fig_plot_ratioyoung_hinged_clamped}, where the behaviour is quite different for the two cases. When the rigid rods are all clamped, $\tilde{E}_{2}/\tilde{E}_{1}=\xi$ independently of $\zeta$. Otherwise, in the hinged case, when $\zeta$ increases, the ratio $\tilde{E}_{2}/\tilde{E}_{1}$ decreases until the minimum at $\zeta=1$ is reached. 

The Poisson's ratios are not null only in the case of hinged connections, because the elastic rods can change their relative internal angles in agreement with the expansion or contraction of the lattice. The extreme values at the boundary of the domain are
\begin{equation}
     \nu_{12} (\zeta=0)= 0 \, , 
     ~~~
     \nu_{21}(\zeta=0) = 0 \, , ~\mbox{ and }~
     \nu_{12}(\zeta=1) = \frac{1}{\xi^{2}} \, , 
     ~~~
     \nu_{21}(\zeta=1) = \xi^{2} \, . 
\end{equation}
Functions $\nu_{12}(\zeta)$ and $\nu_{21}(\zeta)$ are monotonically increasing, so that they 
are maximized at $\zeta=1$. Moreover, the maximum values depend on the aspect ratio $\xi$ and do not present an upper limit. The Poisson's ratios are reported in Fig.~\ref{fig_plot_poisson_hinged} as functions of the parameter $\zeta$.
%
%%%%%%%%%%%%%%%%%%%%%%%%%%%%%%%%%%%%%%%%%%%%%%%%%
%%%%%%%%%%%%%%%%%%%%%%%%%%%%%%%%%%%%%%%%%%%%%%%%%%%%%%%%%%

\begin{figure}
    \centering
    \begin{subfigure}[]{0.48\textwidth}
    \centering
    \includegraphics[width=\textwidth, align=c]{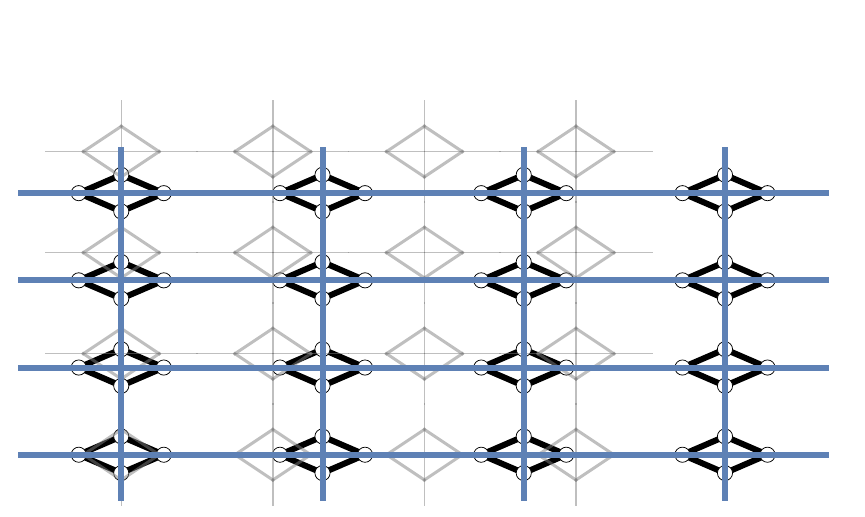}
    \end{subfigure}
    \begin{subfigure}[]{0.48\textwidth}
    \centering
    \includegraphics[width=\textwidth, align = c]{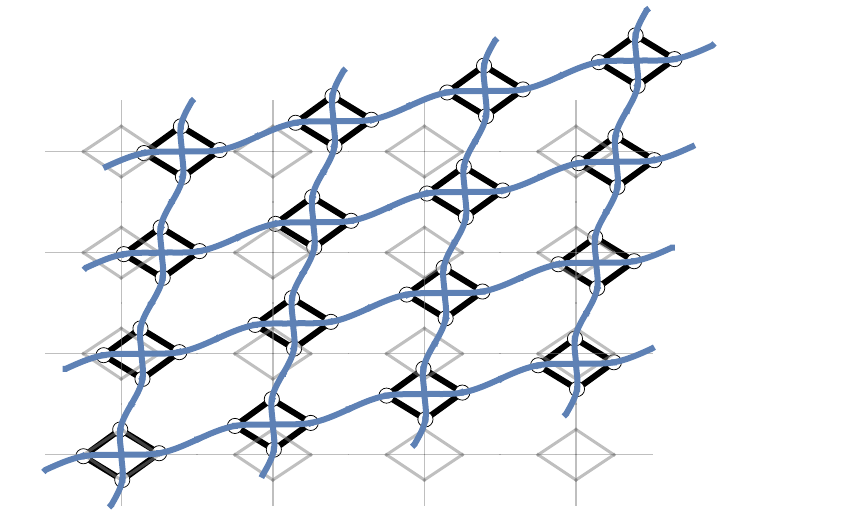}
    \end{subfigure}
    \caption{Uniaxial stress (left) and pure shear (right) of a rectangular lattice with hinged rigid elements, forming rhombic-shaped figures. The uniaxial stress shows a strong Poisson's effect related to the deformation occurring inside the rhombic-shaped elements, while undeformed beams are visible inside the rhombic-shaped elements when the grid is subject to shear. Parameters $\zeta =0.5$  and $\lambda_1=\lambda_2=20$ have been selected.}
    \label{fig_deformed_rectangular_hinged} 
\end{figure}

The lattice with hinged connections is represented in Fig.~\ref{fig_deformed_rectangular_hinged}, when subject to a horizontal uniaxial stress (on the left) and a pure shear stress (on the right). Because of the non-vanishing Poisson's ratios, the axial stress deforms the rhombic-shaped elements. 
In the case of the shear stress, rhombic-shaped elements remain almost undeformed. This behaviour is correlated with the almost flat shape of $\tilde{G}_{12}(\zeta)$ (see Fig.~\ref{fig_plot_shear_hinged_clamped}): since the shearing stiffness is insensitive to the parameter $\zeta$, the hinged rigid rods tend to be not involved in this type of stress application. 
%
%%%%%%%%%%%%%%%%%%%%%%%%%%%%%%%%%%%%%%%%%%%%%%%%%%%%%%%%%%%%%%%
%%%%%%%%%%%%%%%%%%%%%%%%%%%%%%%%%%%%%%%%%%%%%%%%%%%%%%%%%%%%%%%

\section{Conclusions}
\label{Sec_conclusions}

A quasi-static technique for the determination of an elastic continuum equivalent to a periodic grid of elastic beams has been enhanced with the inclusion of rigid elements inside the grid. Even with the presence of the latter elements, the presented homogenization scheme has been shown to lead to the analytic determination of the elasticity tensor. 
Several effects related to the presence of rigid inclusions have been found, including auxeticity and singular behaviour in the rigidity limit. When rigid elements are connected to elastic rods with hinges, it is shown that beams contained inside a zone surrounded by rigid bars can still deform and influence the overall mechanical response. 

The presented formulation makes available analytical expressions for the characterization of elastic solids equivalent to architected materials. Therefore, our results can be used as design tools for new materials based on microstructures.

\section*{Acknowledgements}
\label{sec:funding}

Davide Bigoni remembers with great pleasure his scientific collaboration with Professor Alan Needleman and the happy moments spent together, often with his wife Wanda. He looks forward to continuing to benefit from collaboration with Alan and to enjoy his friendship for many years to come.

All the authors acknowledge financial support from the European Research Council (ERC) under the European Union’s Horizon 2020 research and innovation programme (Grant agreement No. ERC-ADG-2021-101052956-BEYOND).

\appendix

%\section{Appendix}
%\label{sec:app1}
% 

\printbibliography

\end{document}